\documentclass[reprint,amsmath,amssymb,aps,prl,superscriptaddress]{revtex4-2}

\usepackage{graphicx}
\usepackage{dcolumn}
\usepackage{bm}
\usepackage{subcaption}
\usepackage{amsmath}
\usepackage{amssymb}
\usepackage{amsfonts}
\usepackage{color}
\usepackage{bbm}
\makeatletter
\def\widetext@rule{}
\makeatother

\def\be{\begin{equation}}
\def\ee{\end{equation}}
\def\ba{\begin{array}{lll}}
\def\ea{\end{array}}

\def\beq{\begin{eqnarray}}
\def\eeq{\end{eqnarray}}

\newcommand{\genova}{Dipartimento di Fisica, Università di Genova, Via Dodecaneso 33, 16146 Genova, Italy}
\newcommand{\CNR}{CNR-SPIN, Via Dodecaneso 33, 16146 Genova, Italy}
\newcommand{\como}{Center for Nonlinear and Complex Systems, Dipartimento
di Scienza e Alta Tecnologia, Universit\`a degli Studi dell'Insubria,
via Valleggio 11, 22100 Como, Italy}
\newcommand{\infn}{Istituto Nazionale di Fisica Nucleare, Sezione di Milano,
via Celoria 16, 20133 Milano, Italy}

\begin{document}
\title{Quantum advantage bounds for a multipartite Gaussian battery}
\author{F. Cavaliere}
\affiliation{\genova}
\affiliation{\CNR}
\author{D. Ferraro}
\email{dario.ferraro@unige.it}
\affiliation{\genova}
\affiliation{\CNR}
\author{M. Carrega}
\affiliation{\CNR}
\author{G. Benenti}
\affiliation{\como}
\affiliation{\infn}
\author{M. Sassetti}
\affiliation{\genova}
\affiliation{\CNR}



\date{\today}

\begin{abstract}
We demonstrate the possibility of a genuine quantum advantage in the efficiency of quantum batteries by analyzing a model that enables a consistent comparison between quantum and classical regimes. Our system consists of $N$ harmonic oscillator cells coupled to a common thermal reservoir, evolving through Gaussian states. 
We define the global efficiency as the ratio of extractable work (ergotropy) to stored energy, and derive analytical bounds that distinguish, in order of increasing efficiency, regimes characterized by classical squeezing, quantum squeezing without entanglement, and genuine entanglement. Moreover, numerical simulations support the emergence of a similar hierarchy for the thermodynamic efficiency, defined as the ratio between ergotropy and the total thermodynamic cost of the charging process.
\end{abstract}

\maketitle

{\em Introduction}---Inspired by the progressive miniaturization of devices devoted to energy storage~\cite{Dell01}, the concept of Quantum Batteries (QBs) has recently emerged as a hot topic in the panorama of quantum technologies~\cite{Alicki13, Binder15, Bhattacharjee21, Quach23, Campaioli24}. Theoretical proposals and early-stage experimental proofs of principle of devices exploiting quantum features for energy transfer, storage, and manipulation have appeared, based on different platforms. Among them are cavity and circuit quantum electrodynamics~\cite{Ferraro18, Crescente20, Quach22, Shaghaghi22, Carrasco22, Gemme23, Rodriguez23, Erdman24, Wang24, Canzio25, Massa25, Hymas25, Kurman25}, spin chains and networks~\cite{Le18, Rossini20, Catalano24, Grazi24, Grazi25, Donelli25, Beder25}, molecules~\cite{Joshi22, Cruz22}, and harmonic oscillators~\cite{Andolina18, Barra20, Downing23, Konar24, Cavaliere25, Andolina25}. 

Among the main questions emerging in this domain, as well as in all quantum technologies, arguably the most fundamental is:~\emph{``Is there a genuine quantum advantage?''}
As the stored energy $E_B$ scales with the number $N$ of 
elementary units composing the quantum battery, similarly to the classical 
case, research has focused on the charging power $P = E_B / t^*$, where $t^*$ is the charging time. 
A super-extensive scaling, $P \propto N^k$ with $k > 1$, can be achieved via two distinct mechanisms~\cite{Julia20}.
On the one hand, a \emph{collective advantage} can arise, associated with many-body induced accelerations in the QB evolution from the initial empty state to a charged state, which may also occur in classical models~\cite{Ferraro18, Andolina19, Hovhannisyan20}. On the other hand, a \emph{genuine quantum advantage} occurs in the presence of energy fluctuations scaling more than linearly with $N$, and is attributed to entanglement generated among the elementary units composing the QB~\cite{Campaioli17, Gyhm22, Rinaldi24}. 
This latter form of advantage, however, has been demonstrated in models -- e.g., when the QB is composed of $N$ spin-$1/2$ units -- that lack a classical analog to compare with. 
More importantly, in analogy with quantum thermal machines~\cite{Vinjanampathy2016,Benenti2017,Daffner2019, Arrachea23, Cangemi24,Razzoli25}, a proper thermodynamic quantification of the performance of a QB must take into account the \emph{efficiency} of work extraction. The possibility of a quantum advantage in this context, however, has so far eluded investigation.

The purpose of this Letter is to address the possibility of a \emph{quantum advantage in efficiency} by considering a system that can be treated both in a fully quantum regime and in its semi-classical limit, thereby allowing for a fair comparison of the QB performance across different regimes. That is, we compare configurations with the same number of cells, stored energy, and temperature.
Specifically, we consider a QB composed of $N$ identical harmonic oscillator cells, 
initially prepared at the same temperature and coupled to a common reservoir. 
Throughout the entire charging process, the state of the QB remains Gaussian~\cite{Braunstein05, Adesso07, Adesso14}, and dynamics is characterized by the generation of squeezing~\cite{Cavaliere25}.
By considering the \emph{global efficiency} $\eta_{\mathrm{glob}}(t)$, 
defined as the ratio between the work extractable via global unitary operations acting on the $N$ cells (\emph{ergotropy}) and the stored energy, we derive rigorous bounds
establishing a clear quantum advantage:
$\eta_{\mathrm{glob}}(t)< \mathcal{B}_{\mathrm{cl}}(t)$
in the case of classical squeezing;
$\mathcal{B}_{\mathrm{cl}}(t)< \eta_{\mathrm{glob}}(t)\le \mathcal{B}_{\mathrm{en}}(t)$
with quantum squeezing without entanglement; and 
$\mathcal{B}_{\mathrm{en}}(t)<\eta_{\mathrm{glob}}(t)\le 1$
when entanglement is present.
Finally, numerical results support the existence of the same hierarchy 
of performance for the \emph{thermodynamic efficiency} $\eta_{\mathrm{th}}(t)$, 
defined as the ratio between the ergotropy and the overall thermodynamic cost 
of the charging process.

{\em Model and covariance matrix}---The total Hamiltonian   is given by 
$H_{\mathrm{tot}}=H_{\mathrm{B}}+H_{\mathrm{int}}+H_{\mathrm{R}}$,
with 
\begin{equation}
H_{\mathrm{B}}=\sum^{N}_{l=1}\left(\frac{p^{2}_{l}}{2m}+\frac{m \omega_{0}^2}{2} q^{2}_{l}\right)=\sum^{N}_{l=1}H^{(l)}_{\mathrm{B}}
\label{eq:HB}
\end{equation}
the Hamiltonian of the QB, composed by $N$ \emph{identical} harmonic oscillators with equal frequencies and masses.
\noindent The reservoir is modeled by a set of harmonic oscillators~\cite{Ingold02, Weiss12, Freitas14, Carrega16, Carrega20, Carrega22}, 
$H_{\mathrm{R}}=\sum_{k=1}^{+\infty}\left(\frac{\pi^{2}_{k}}{2m_k}+\frac{m_k \omega_{k}^2 x^{2}_{k}}{2}\right)$, interacting with the QB through the coupling~\cite{Carrega24} 
\begin{equation}
\label{eq:Hint}
\!\!\!\!H_{\mathrm{int}}=-\sum^{+\infty}_{k=1}c_{k}x_{k} \sum^{N}_{l=1} \alpha_{l} q_{l}+\sum^{+\infty}_{k=1}\frac{c^{2}_{k}}{2 m_{k} \omega^{2}_{k}}\left(\sum^{N}_{l=1} \alpha_{l} q_{l} \right)^{2}\,
\end{equation}
which is turned on at time $t=0$.
\\
Notice that the coupling with the $l$-th oscillator of the QB is
modulated  by the 
factor $\alpha_{l}$. The initial state is described by the density matrix $\rho_{\mathrm{tot}}(0)=\frac{e^{-H_{\mathrm{B}}/k_{\mathrm{B}} T_0}}{Z_{\mathrm{B}}} \otimes  \frac{e^{-H_{\mathrm{R}}/k_{\mathrm{B}} T}}{Z_{\mathrm{R}}}$, 
the tensor product of two initially 
uncorrelated thermal states, with $T_{0}$ the temperature of the QB, and $T$ the temperature of the reservoir.


The dynamical properties of the QB are characterized by the time behavior of its covariance matrix (CM)~\cite{Serafini2017,Brask2022}.
Rather than adopting the usual \emph{local} representation, expressed in terms of the canonical quadratures $Q_{l}=\sqrt{\frac{m \omega_{0}}{\hbar}}q_{l}$ and $
P_{l}=\sqrt{\frac{1}{\hbar m \omega_{0} }}p_{l}$ (see~\cite{SM}), 
it is useful to switch to a \emph{global} basis, which highlights the \emph{collective} properties of the model.
%
Indeed, Eq.~(\ref{eq:Hint}) shows that only the combination $\sum_{l}\alpha_lQ_l(t)$ couples to the reservoir. This observation allows us to introduce a {\em bright mode} (BM)~\cite{Ciuti05}, with quadratures
\begin{equation}
\label{eq:BM}
(Q_{\mathrm{BM}},P_{\mathrm{BM}})\equiv\frac{1}{\bar{\alpha}} \sum^{N}_{l=1}\alpha_{l}(Q_{l},P_{l});\quad
\bar{\alpha}=\sqrt{\sum_{l=1}^N\alpha_l^2},
\end{equation}
which is the only mode that couples to the reservoir.
The remaining $N-1$ are {\em dark modes} (DMs), which are completely decoupled from the reservoir. Each DM has energy frozen at the initial value $\frac{\hbar\omega_0}{2}C(T_0)$, with
$C(T_{0})=\coth\left(\frac{\hbar\omega_0}{2k_{\mathrm{B}}\,T_{0}}\right)$.
On this global basis, the covariance matrix of the QB,
  ${{\sigma}}_{\mathrm{B}}^{(\mathrm{glob})}(t)$, takes a block--diagonal form~\cite{SM}
  with a single block corresponding to the BM:
\begin{equation}
\label{eq:SigmaBM}
{{\Sigma}}_{\mathrm{BM}}(t)=\begin{pmatrix}
\frac{C(T_0)}{2}+\bar{\alpha}^2a(t)&\bar{\alpha}^2b(t)\\
\bar\alpha^2b(t)&\frac{C(T_0)}{2}+\bar{\alpha}^2c(t)
\end{pmatrix}\,,    
\end{equation}
where the expressions for $a(t)$, $b(t)$, and $c(t)$ are provided in~\cite{SM}.
The $N-1$ DMs are identical and represented by blocks of the form
$\Sigma_{\mathrm{DM}}=\frac{1}{2}\mathrm{diag}\left\{C(T_0),C(T_0)\right\}$.\\
\begin{figure}[ht]
\includegraphics[width=0.95\columnwidth]{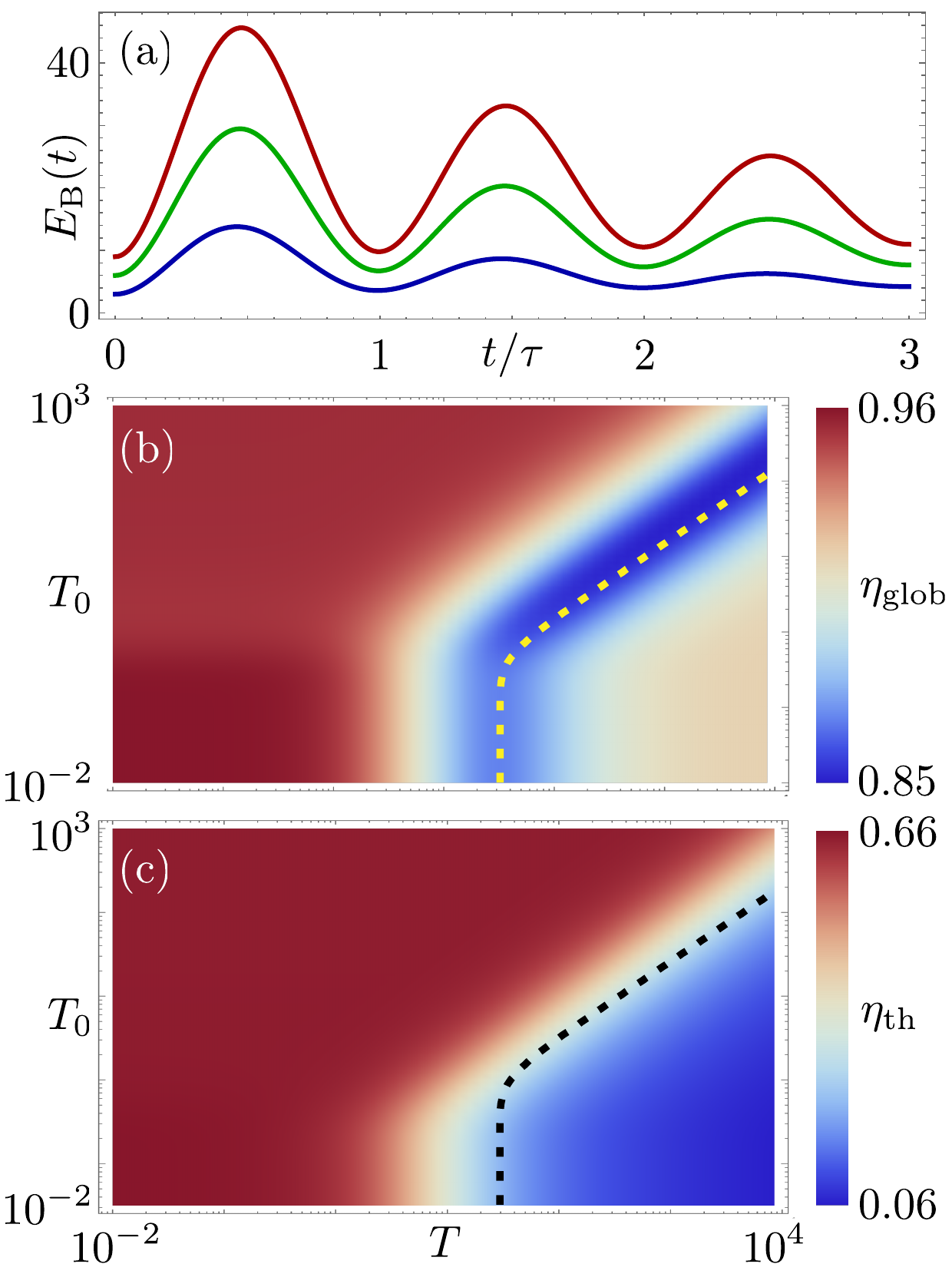}
\caption{(a) Plot of $E_{\mathrm{B}}(t)$ (units $\hbar\omega_0$) as a function of $t/\tau$ for $N=6$ (blue), $N=12$ (green) and $N=18$ (red). For $N=6$, density plots of $\eta_{\mathrm{glob}}=\eta_{\mathrm{glob}}(t^*)$ (b) and $\eta_{\mathrm{th}}=\eta_{\mathrm{th}}(t^*)$ (c),  as a function of $T$ and $T_0$ (units $\hbar\omega_0/k_{\mathrm{B}}$). The dashed line in panels (b) and (c) represents the boundary $T=T^{*}$. Here $\alpha_l=1$, $\gamma_0=5\,\omega_0$ and $\omega_D=2\,\omega_0$.}
\label{fig:Fig1}
\end{figure}
{\em Energy storage and extraction}---Given the above structure of the covariance matrix, the total energy of the battery at time $t$, defined as  $E_{\mathrm{B}}(t)\equiv\langle H_B(t)\rangle$, can be written as
\begin{equation}
\label{eq:Etot}
E_{\mathrm{B}}(t)=E_{\mathrm{BM}}(t)+(N-1)\frac{\hbar\omega_0}{2}C(T_0)\,,
\end{equation}
where 
$
E_{\mathrm{BM}}(t)=\frac{\hbar\omega_0}{2}\left\{C(T_0)+\bar{\alpha}^2\big[a(t)+c(t)\big]\right\}$
is the energy associated with the BM, obtained 
by taking the trace 
in Eq.~(\ref{eq:SigmaBM}).
The behavior of $E_{\mathrm{B}}(t)$ is shown in Fig.~\ref{fig:Fig1}(a) for three different values of $N$, plotted as a function of the rescaled time $t/\tau$, where $\tau = \pi/\Omega_N$. The characteristic frequency $\Omega_N = \bar{\alpha} \sqrt{\gamma_0 \omega_{\mathrm{D}}}$ defines the typical time scale of the battery dynamics, and is determined by the coupling strength to the reservoir $\gamma_0$ and the cutoff frequency $\omega_{\mathrm{D}}$ of the spectral density~\cite{SM}.
The stored energy exhibits oscillations with several local maxima, 
the largest of which occurs at $t = t^* \approx \tau/2$. 
We note that $E_{\mathrm{B}}(t^*)\propto N$, and the charging power $P(t^*)=E_{\mathrm{B}}(t^*)/t^*\propto N^{3/2}$. 
Such super-extensive scaling is analogous to that observed in other systems exhibiting a {\em collective advantage}~\cite{Ferraro18, Crescente20, Barra20, Quach22, Hymas25}.

In addition to energy, a key quantity is the {\em ergotropy}~\cite{Allahverdyan04}.
For a Gaussian system, the ergotropy is directly related to the symplectic eigenvalues of the CM~\cite{Farina19, Downing23}.
By considering all possible \emph{global unitary} operations acting on the $N$-oscillator system, we obtain the global ergotropy
$\mathcal{E}_{\mathrm{glob}}(t)=E_{\mathrm{B}}(t)-\hbar\omega_0\sum_{j=1}^N\nu_j(t)$, where $\nu_j(t)$ are the symplectic eigenvalues of 
${{\sigma}}_{\mathrm{B}}^{(\mathrm{glob})}(t)$, with 
\begin{equation}
\label{eq:nu1}
\nu_{1}(t)=\frac{1}{2}\sqrt{C^2(T_0)+2\bar{\alpha}^{2}\left[ C(T_0)\,\mathcal{T}(t)+2 \bar{\alpha}^{2} 
\Delta(t)\right]}\,,
\end{equation}
stemming from the BM  and $\nu_{2\le j\le N}=C(T_0)/2$ from the DMs. Here, $\mathcal{T}(t)=a(t)+c(t)$, $\Delta(t)=a(t)c(t)-b^2(t)$. Then,
\begin{equation}
\mathcal{E}_{\mathrm{glob}}(t)=E_{\mathrm{BM}}(t)-\hbar\omega_0\nu_1(t)\,.   
\label{eglob}
\end{equation}

To assess the performance of the QB, we consider two definitions of efficiency. The \emph{global efficiency} is the ratio between the global 
ergotropy and the energy stored in the battery~\cite{Barra22} :
\begin{equation}
\label{eq:etaGlob}
\eta_{\mathrm{glob}}(t)=\frac{\mathcal{E}_{\mathrm{glob}}(t)}{E'_{\mathrm{BM}}(t)}=1-\frac{\hbar\omega_0}{E'_{\mathrm{BM}}(t)}[\nu_1(t)-1/2]\,,
\end{equation}
where the energy $E'_{\mathrm{BM}}$ is obtained from $E_{\mathrm{BM}}$
by subtracting the (unextractable) zero-point energy 
$\hbar\omega_0/2$.
The density plot of $\eta_{\mathrm{glob}}(t^*)$ in Fig.~\ref{fig:Fig1}(b) reveals the emergence of two distinct regions, separated approximately by a characteristic temperature $T^*=\frac{\hbar\Omega_N^2}{2k_{\mathrm{B}}T}\coth\left(\frac{\hbar\omega_0}{2k_{B}T_0}\right)$,
which results from the interplay between thermal fluctuations in the reservoir and those in the initial state of the QB.
When $T > T^*$, the reservoir thermal fluctuations dominate. These fluctuations are not useful for enhancing global work extraction, as they mainly generate classical correlations. 
Conversely, when $T < T^*$, thermal fluctuations are suppressed, enabling global unitary operations to extract more work than \emph{local} operations applied independently to each oscillator (see~\cite{SM}).


The second efficiency we consider, termed \emph{thermodynamic efficiency},
is defined as the ratio between the global ergotropy and the total thermodynamic cost associated with the charging process.
To that end, we make use of the concepts of entropy production, $\Sigma_{irr}(t)$~\cite{Esposito10, Rao16, Landi21, Aguilar25}, 
and exergy $\Phi(t)$~\cite{Manzano20, Lu23, Lopez23}.
The latter is defined as the ratio between the negative and positive contributions to the entropy production:
$\Phi(t)=-\Sigma_{\rm irr}^{(-)}(t)/\Sigma_{\rm irr}^{(+)}(t)$.
Due to the second law of thermodynamics, which imposes $\Sigma_{\rm irr}(t)\ge 0$, the exergy is constrained to the range $0\leq\Phi(t)\leq 1$ ~\cite{Manzano20, Lu23, Lopez23, Cavaliere23, Finocchiaro25}. 
As shown in the End Matter, $\Phi(t)$ admits a closed-form expression, from which the thermodynamic efficiency can be derived:
\begin{equation}
\label{eq:etaTh}
\!\!\!\!\!\!\eta_{\mathrm{th}}(t)\!=\!\frac{\mathcal{E}_{\mathrm{glob}}(t)}{\!W(t)\!+\!T\theta[\Delta S_{\mathrm{B}}(t)]\Delta S_{\mathrm{B}}(t)\!+\!\theta[\Delta\epsilon(t)]\Delta\epsilon(t)}.
\end{equation}
Here, $W(t)\!=\!\langle H_{\mathrm{int}}(0) \rangle\!-\! \langle\!H_{\mathrm{int}}(t)\rangle$ is the connection/disconnection work~\cite{SM}, $\Delta S_{\mathrm{B}}(t)=S_{\mathrm{B}}(t)-S_{\mathrm{B}}(0)$ the Von Neumann entropy cost of the battery, $\Delta\epsilon(t)=\hbar\omega_0\left[C(T_0)/2-\nu_1(t)\right]$ and $\theta(x)$ the Heaviside step function.  
Note that the thermodynamic efficiency is always bounded as
$0\!\le\!\eta_{\mathrm{th}} (t)\leq\Phi(t)\leq\! 1$ (see End Matter).
The efficiency $\eta_{\mathrm{th}}(t^*)$ is shown in Fig.~\ref{fig:Fig1}(c). It confirms that the regime $T < T^*$ is the most favorable for the QB, while for $T > T^*$ strong thermal fluctuations impose a significant entropic cost, resulting in a reduction of $\eta_{\mathrm{th}}(t)$.


{\em Quantum advantage}---From the observations made so far, it is evident that the best overall performance is achieved in the regime $T < T^*$, where reservoir fluctuations are weak. Interestingly, in this same regime, high values of $T_0$--corresponding to a nearly classical battery--can also lead to good performance; see Figs.~\ref{fig:Fig1}(b,c).
This raises two important questions: Are there signatures of {\em genuine quantum correlations}? And if so, what role do they play in determining the battery’s performance?
Addressing these questions is the central goal of this work.

To begin, we investigate the structure of the BM, which corresponds to a (phase--space rotated) {\em squeezed thermal state}~\cite{Brask2022,Serafini2017}. 
In the rotated frame its CM, ${{\Sigma}}'_{\mathrm{BM}}(t)$, is diagonal and consists of the variances of the canonical quadratures $Q'_{\mathrm{BM}}$,~$P'_{\mathrm{BM}}$ along the principal axes. The corresponding eigenvalues are (see End Matter)
\begin{equation}
\label{eq:eigenvaluespm}
\!\!\!\!\!\lambda_{\pm}(t)=\frac{1}{2}\left\{C(T_0)+\bar{\alpha}^2\!\left[\mathcal{T}(t)\pm\sqrt{\mathcal{T}^2(t)-4\Delta(t)}\right]\right\}\!\,,
\end{equation}
satisfying $\lambda_{+}(t)\lambda_{-}(t)\geq 1/4$ always. They define the {\em squeezing parameter}  $r(t)=\frac{1}{4}\ln\left[\lambda_+(t)/\lambda_-(t)\right]$, here written for $r(t)>0$. 
Note also that
\begin{equation}
\!\!\!\!\!E_{\mathrm{BM}}(t)\!=\!\frac{\hbar\omega_0}{2}[\lambda_{+}(t)+\lambda_{-}(t)].\label{eq:relationships}
\end{equation}
An important distinction must be made regarding the type of squeezing. When both eigenvalues satisfy $\lambda_{\pm}(t) > 1/2$, the BM is in a {\rm classical squeezed state}, where the fluctuations of $Q'_{\mathrm{BM}}$ and $P'_{\mathrm{BM}}$ have a semiclassical origin. In contrast, when $\lambda_{-}(t)<1/2$, one of the variances drops {\em below the ground-state value}, signaling that the BM enters a regime of {\em genuine quantum squeezing}~\cite{Serafini2017,Brask2022}.

\begin{figure}[h!t]
    \centering
    \includegraphics[width=0.45\textwidth]{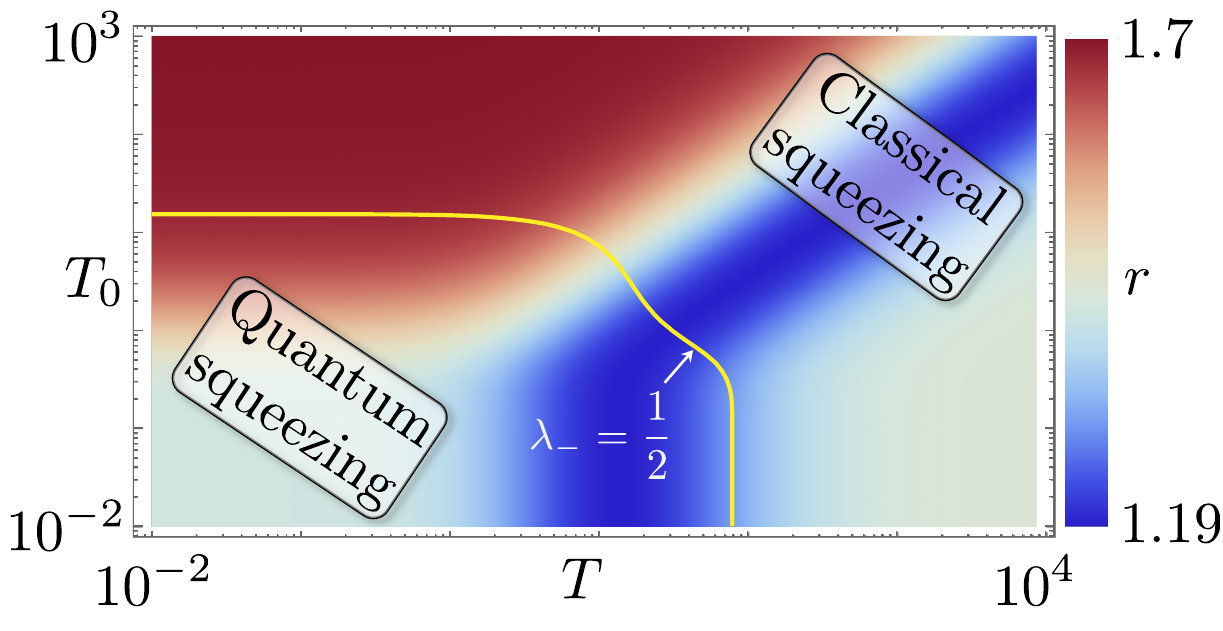}
    \caption{Density plot of $r=r(t^*)$ as a function of $T$ and $T_0$ (units $\hbar\omega_0/k_{\mathrm{B}}$). The yellow  line corresponds to $\lambda_-(t^*)=1/2$. Here $N\!=\!4$, $\alpha_l\!=\!1$, $\gamma_0\!=\!5\,\omega_0$, $\omega_D=2\,\omega_0$.}
    \label{fig:Fig3}
\end{figure}

Figure~\ref{fig:Fig3} displays the regions of classical and quantum squeezing in the $(T, T_0)$ parameter space. These regions are separated by the yellow contour line corresponding to $\lambda_{-}(t^*) = 1/2$. The underlying density plot shows the squeezing parameter $r(t^*)$, revealing that the BM is {\em always squeezed}. As both $T$ and $T_0$ decrease, the system transitions into a broad regime of {\em quantum squeezing}. This high degree of controllability and the ability to access distinct squeezing regimes make this setup an ideal candidate to quantitatively analyze the presence of genuinely quantum advantage.

To proceed, we must first establish a fair comparison between classical and quantum regimes. For this purpose, we consider two configurations with the same number of oscillators $N$, identical initial and reservoir temperatures $T_0$ and $T$, and equal energy $E_{\mathrm{B}}(t^*)$ stored in the battery. Tuning the coupling strength $\gamma_0$ and the spectral cutoff frequency $\omega_{\mathrm{D}}$,  it is possible to access either the classical or the quantum squeezing regime.

Under the above conditions, we demonstrate the existence of a bound $\mathcal{B}_{\mathrm{cl}}(t)\leq 1$ on $\eta_{\mathrm{glob}}(t)$ such that
\begin{eqnarray}
&&\hspace{0.6cm}0\leq\eta_{\mathrm{glob}}(t)<\mathcal{B}_{\mathrm{cl}}(t)\label{bound1}\nonumber\\
&&\hspace{-0.85cm}{\rm with\, classical \;\,squeezing}\;\, (\lambda_-(t)>1/2),
\nonumber\\&&\nonumber\\
&&\hspace{0.6cm}\mathcal{B}_{\mathrm{cl}}(t)<\eta_{\mathrm{glob}}(t)\leq1\label{bound2}\nonumber\\
&&\hspace{-0.85cm}{\rm with\, quantum \,\,squeezing}\;\, (\lambda_-(t)<1/2),
\nonumber
\end{eqnarray}
where (see End Matter):
\begin{equation}
\label{eq:bound1}
\mathcal{B}_{cl}(t)=1-\frac{1}{2\kappa(t)}\left[\sqrt{4\kappa(t)+1}-1\right],
\end{equation}
with $\kappa(t)=\frac{E'_{\mathrm{BM}}(t)}{\hbar\omega_0}>0$. This proves that {\em quantum squeezing increases the performances of the  battery} with respect to a classically squeezed regime.

\begin{figure}[h!t]
    \centering
    \includegraphics[width=0.475\textwidth]{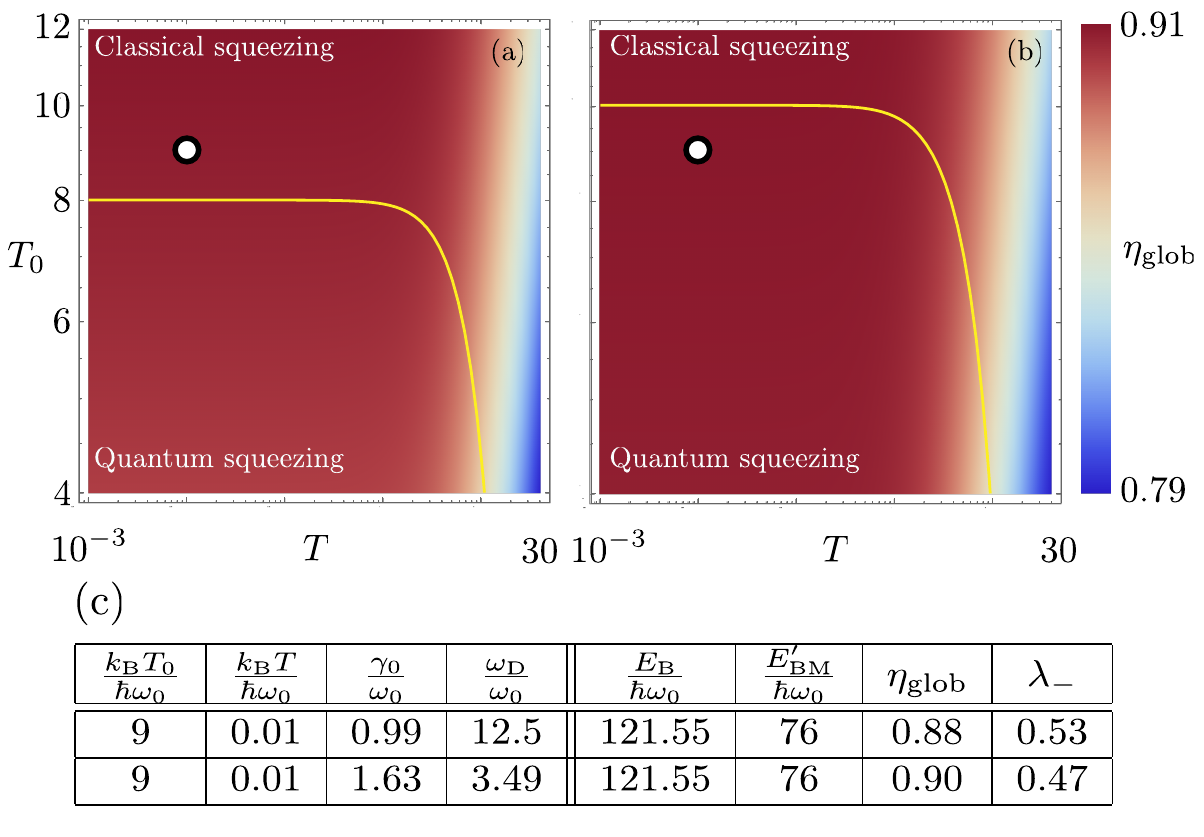}
    \caption{Density plot of $\eta_{\mathrm{glob}}=\eta_{\mathrm{glob}}(t^*)$ as a function of $T$ and $T_0$ (units $\hbar\omega_0/k_{\mathrm{B}}$) with $N=6$ and (a) $\gamma_0=0.99\,\omega_0$, $\omega_D=12.5\,\omega_0$; (b)  $\gamma_0=1.63\,\omega_0$, $\omega_D=3.49\,\omega_0$. 
    In both panels, the white dot indicates the common operating point chosen for the comparison, while the yellow line denotes the boundary between classical and quantum squeezing, defined by the condition $\lambda_-(t^*)=1/2$. 
    In (a), the operating point lies in the classical squeezing region, while in (b) it falls within the quantum squeezing regime.
    (c) Table summarizing the precise values of the model parameters and relevant physical quantities corresponding to the operating point in both scenarios.}
    \label{fig:Fig4}
\end{figure}

Figure~\ref{fig:Fig4} illustrates the advantages of quantum squeezing in a specific case by showing density plots of $\eta_{\mathrm{glob}}(t^*)$ for a system with $N=6$ oscillators. 
The white dot in panels (a) and (b) marks the chosen operating point.
By varying the coupling strength $\gamma_0$ and the spectral cutoff $\omega_D$ between the two panels, the position of the boundary defined by $\lambda_-(t^*) = 1/2$ (yellow line), which separates classical and quantum squeezing regimes, also changes.
As a result, the same operating point lies within the classical squeezing regime in panel (a), and within the quantum squeezing regime in panel (b).
The exact parameter values used in each case are summarized in Table~(c). 

Having established a clear quantum advantage arising from quantum squeezing, we now turn our attention to a qualitatively different quantum feature: {\em entanglement~}\cite{Horodecki09}. Specifically, we investigate the emergence and role of entanglement by considering a balanced bipartition of the $N$ oscillators into two subsystems, $A$ and $B$, each containing $N_{\mathrm{A}} = N_{\mathrm{B}} = N/2$ oscillators (assuming even $N$), with balanced couplings $\sum_{l=1}^{N_A}\alpha^2_l=\sum_{l=N_A+1}^{N}\alpha^2_l=\bar\alpha^2/2$ (details in End Matter). 

As an entanglement witness, we adopt the logarithmic negativity~\cite{Serafini2017}, which, in our case, is expressed in terms of the smallest symplectic eigenvalue $\nu_-^{(\mathrm{PT})}(t)$ of the partially transposed covariance matrix (CM) associated with the balanced bipartition of the system (see End Matter for details). It is defined as
\begin{equation}
\label{eq:logneg}
\mathcal{N}(t)=\mathrm{max}\left\{0,-\ln\left[2\nu_-^{(\mathrm{PT})}(t)\right]\right\}\,.
\end{equation}
Entanglement is present when $\mathcal{N}(t) > 0$, that is, when $\nu_-^{(\mathrm{PT})}(t) < 1/2$. Interestingly, there exists a strong connection between $\nu_-^{(\mathrm{PT})}(t)$ and the minimal eigenvalue $\lambda_-(t)$ of the BM covariance matrix: in fact, we find (see End Matter)
\begin{equation}
\label{eq:result1}
\nu_-^{(\mathrm{PT})}(t)=\sqrt{\frac{C(T_0)}{2}}\sqrt{\lambda_{-}(t)}\,.
\end{equation}
Since $C(T_0)\ge 1$, 
we conclude that {\em entanglement implies quantum squeezing}, while the converse does not necessarily hold.

\begin{figure}[h!t]
    \centering
    \includegraphics[width=0.45\textwidth]{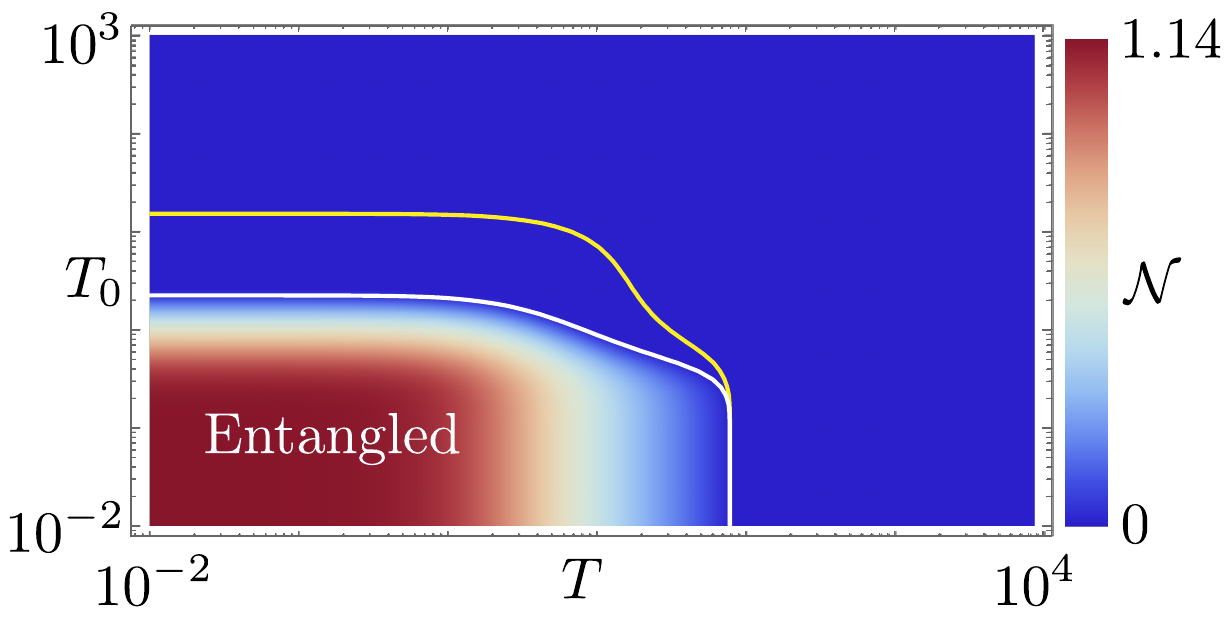}
    \caption{Density plot of $\mathcal{N}=\mathcal{N}(t^*)$ as a function of $T$ and $T_0$ (units $\hbar\omega_0/k_{\mathrm{B}})$. The white line, $\nu_-^{(\mathrm{PT})}(t^*)=1/2$, marks the boundary between entangled and separable states. The yellow line, $\lambda_-(t^*)=1/2$, separates the regions of quantum and classical squeezing (see Fig.~\ref{fig:Fig3}).
    Here, $N=4$, with a balanced partition $N_{\mathrm{A}}\!=\!N_{\mathrm{B}}\!=\!2$, $\alpha_l=1$, $\gamma_0=5\,\omega_0$ and $\omega_{\mathrm{D}}=2\,\omega_0$.}
    \label{fig:Fig5}
\end{figure}

Figure~\ref{fig:Fig5} shows the typical behavior of $\mathcal{N}=\mathcal{N}(t^*)$ as a function of  $T$ and $T_0$. As expected entanglement emerges at low $T$ and $T_0$, within the region bounded by the white line ($\nu_-^{(\mathrm{PT})}(t^*)=1/2$).
This region lies entirely inside the yellow boundary of quantum squeezing
($\lambda_-(t^*)=1/2$), thus confirming that entanglement implies quantum squeezing. 

We are now in a position to investigate another source of quantum advantage: the role of entanglement. Specifically, we compare two scenarios --- one in which entanglement is present, and another where only quantum squeezing occurs, with no entanglement. 
The comparison is made under fixed conditions: same number of oscillators $N$, same temperatures $T_0$ and $T$, and equal energy stored in the battery $E_{\mathrm{B}}(t)$. By tuning the coupling strength $\gamma_0$ and the spectral cutoff frequency $\omega_D$, we can move between an entangled and a non-entangled regime while keeping these constraints.
Under these conditions, we demonstrate the existence of a new bound, $\mathcal{B}_{\mathrm{en}}(t)$, on the efficiency $\eta_{\mathrm{glob}}(t)$, which leads to a \emph{distinct form of quantum advantage}, summarized as follows:
\begin{eqnarray}
&&\hspace{0.5cm}\mathcal{B}_{\mathrm{cl}}(t)<\eta_{\mathrm{glob}}(t)\le\mathcal{B}_{\mathrm{en}}(t)\nonumber\\
&&\hspace{-1cm}{\rm with\, quantum \,\,squeezing}\,{\rm but\,no\, entanglement},\nonumber\\&&\nonumber\\
&&\hspace{1cm}\mathcal{B}_{\mathrm{en}}(t)<\eta_{\mathrm{glob}}(t)\leq1\nonumber\\
&&\hspace{1.2cm}{\rm with\, entanglement}.\nonumber
\end{eqnarray}
These new bounds prove that {\em entanglement increases the performances of the QB} .
Explicitly (see End Matter):
\begin{equation}
\label{eq:bound2}
\mathcal{B}_{\rm en}(t)=1-\frac{1}{2\kappa(t)}\left[\sqrt{\left[4\kappa(t)+2-\frac{1}{C(T_0)}\right]\frac{1}{C(T_0)}}-1\right]. 
\end{equation}
Note that $\mathcal{B}_{\rm cl}(t)<\mathcal{B}_{\rm en}(t)\leq 1$ always.
\begin{table}[ht]
\centering
\begin{tabular}{|c|c|c|c||c|c|c|c|c|}
\hline
$\frac{k_{\mathrm{B}}T_0}{\hbar\omega_0}$ & $\frac{k_{\mathrm{B}}T}{\hbar\omega_0}$ & $\frac{\gamma_0}{\omega_0}$ & $\frac{\omega_{\mathrm{D}}}{\omega_0}$ & $\frac{E_{\mathrm{B}}}{\hbar\omega_0}$ & $\frac{E'_{\mathrm{BM}}}{\hbar\omega_0}$ & $\eta_{\mathrm{glob}}$ & $\lambda_{-}$ & $\mathcal{N}$ \\
\hline\hline
1&0.01&0.528&7.344&8.28&4.5& 0.72 & 0.29 &0\\
\hline
1&0.01&1.571& 0.922& 8.28 &4.5& 0.79 & 0.19 &0.1\\
\hline
\end{tabular}
\caption{Parameters and results for a battery with $N=6$, $k_{\mathrm{B}}T_0=\hbar\omega_0$, $k_{\mathrm{B}}T=0.01\,\hbar\omega_0$ and $E_{\mathrm{B}}(t^*)=8.28\,\hbar\omega_0$, without or with the presence of entanglement. Here, $N_{\mathrm{A}}=N_{\mathrm{B}}=3$ and $\!\alpha_l=1$.}
\label{tab:Tab1}
\end{table}
\noindent Table~\ref{tab:Tab1} illustrates a concrete example: 
$E_{\mathrm{B}}(t^*)=8.28\,\hbar\omega_0$, 
for two different $\gamma_0$ and $\omega_D$,
entanglement is either present (${\cal N}(t^*)>0$) or absent (${\cal N}(t^*)=0$). Clearly, when entanglement is present, the efficiency $\eta_{\mathrm{glob}}(t^*)$ is larger.

We conclude by reporting numerical evidence suggesting that the thermodynamic efficiency $\eta_{\mathrm{th}}(t^*)$ follows similar trends to the global efficiency $\eta_{\mathrm{glob}}(t^*)$.
Indeed, for the comparison equivalent to Fig.~\ref{fig:Fig4} we  find $\eta_{\mathrm{th}}(t^*)=0.14$ in the classical squeezing regime and  $\eta_{\mathrm{th}}(t^*)=0.38$ in the quantum squeezing regime. Regarding entanglement, as in Tab.~\ref{tab:Tab1}, we obtain $\eta_{\mathrm{th}}(t^*)=0.14$ in the absence of entanglement, 
and $\eta_{\mathrm{th}}(t^*)=0.58$ when entanglement is present.

{\em Conclusions}---We have demonstrated that multipartite Gaussian 
QBs can exhibit a genuine quantum advantage in 
their global efficiency, with analytically derived bounds establishing a clear hierarchy from classical squeezing to quantum squeezing and entanglement.
Finding analogous analytical bounds for the thermodynamic efficiency, 
which accounts for the overall thermodynamic cost of a QB, remains 
an interesting open question.
The proposed model can be realized in the context of quantum $LC$ circuits~\cite{Vool17}, embedded in a dissipative environment with tunable dissipation strength~\cite{Blais21, Vool17, Cattaneo21}. This generalizes the analysis in Ref.~\cite{Cavaliere25} from a single $LC$ quantum battery to a series of $N$ coupled elements. Indeed, each single cell composing the Gaussian QB can be thought of as a single $LC$ element, where the capacitance $C$ plays the role of the mass $m$, and the inductance maps as $L \to 1/(m\omega_0^2)$.
The associated circuit energy is the sum of the energies of all individual $LC$ cells and corresponds to the QB Hamiltonian given in Eq.~(\ref{eq:HB}). 
The dissipative environment can be modeled using a  lumped-element model~\cite{Gramich11, Cavaliere25}. 
In this framework, the dissipation strength is inversely proportional to the environment resistance, i.e., $\gamma_0 \propto R_{\rm env}^{-1}$.

\begin{acknowledgements} G.B, F.C. and D.F. acknowledge support from the project PRIN 2022 - 2022XK5CPX (PE3) SoS-QuBa - "Solid State Quantum Batteries: Characterization and Optimization". M.C. and M.S. acknowledge support from the project PRIN 2022 - 2022PH852L (PE3) - "Non reciprocal supercurrent and Topological Transition in hybrid Nb-InSb nanoflags", both projects  funded within the programme "PNRR Missione 4 - Componente 2 - Investimento 1.1 Fondo per il Programma Nazionale di Ricerca e Progetti di Rilevante Interesse Nazionale (PRIN)", funded by the European Union - Next Generation EU". G.B. acknowledges support from INFN through the project QUANTUM.
\end{acknowledgements}

\bibliography{references}

\begin{thebibliography}{79}%
\makeatletter
\providecommand \@ifxundefined [1]{%
 \@ifx{#1\undefined}
}%
\providecommand \@ifnum [1]{%
 \ifnum #1\expandafter \@firstoftwo
 \else \expandafter \@secondoftwo
 \fi
}%
\providecommand \@ifx [1]{%
 \ifx #1\expandafter \@firstoftwo
 \else \expandafter \@secondoftwo
 \fi
}%
\providecommand \natexlab [1]{#1}%
\providecommand \enquote  [1]{``#1''}%
\providecommand \bibnamefont  [1]{#1}%
\providecommand \bibfnamefont [1]{#1}%
\providecommand \citenamefont [1]{#1}%
\providecommand \href@noop [0]{\@secondoftwo}%
\providecommand \href [0]{\begingroup \@sanitize@url \@href}%
\providecommand \@href[1]{\@@startlink{#1}\@@href}%
\providecommand \@@href[1]{\endgroup#1\@@endlink}%
\providecommand \@sanitize@url [0]{\catcode `\\12\catcode `\$12\catcode `\&12\catcode `\#12\catcode `\^12\catcode `\_12\catcode `\%12\relax}%
\providecommand \@@startlink[1]{}%
\providecommand \@@endlink[0]{}%
\providecommand \url  [0]{\begingroup\@sanitize@url \@url }%
\providecommand \@url [1]{\endgroup\@href {#1}{\urlprefix }}%
\providecommand \urlprefix  [0]{URL }%
\providecommand \Eprint [0]{\href }%
\providecommand \doibase [0]{https://doi.org/}%
\providecommand \selectlanguage [0]{\@gobble}%
\providecommand \bibinfo  [0]{\@secondoftwo}%
\providecommand \bibfield  [0]{\@secondoftwo}%
\providecommand \translation [1]{[#1]}%
\providecommand \BibitemOpen [0]{}%
\providecommand \bibitemStop [0]{}%
\providecommand \bibitemNoStop [0]{.\EOS\space}%
\providecommand \EOS [0]{\spacefactor3000\relax}%
\providecommand \BibitemShut  [1]{\csname bibitem#1\endcsname}%
\let\auto@bib@innerbib\@empty
\bibitem [{\citenamefont {Dell}\ \emph {et~al.}(2001)\citenamefont {Dell}, \citenamefont {Rand}, \citenamefont {Connor},\ and\ \citenamefont {Bailey}}]{Dell01}%
  \BibitemOpen
  \bibfield  {author} {\bibinfo {author} {\bibfnamefont {R.~M.}\ \bibnamefont {Dell}}, \bibinfo {author} {\bibfnamefont {D.~A.~J.}\ \bibnamefont {Rand}}, \bibinfo {author} {\bibfnamefont {P.}~\bibnamefont {Connor}},\ and\ \bibinfo {author} {\bibfnamefont {R.~B.~D.}\ \bibnamefont {Bailey}},\ }\href {https://doi.org/10.1039/9781847552228} {\emph {\bibinfo {title} {Understanding Batteries}}}\ (\bibinfo  {publisher} {The Royal Society of Chemistry},\ \bibinfo {year} {2001})\BibitemShut {NoStop}%
\bibitem [{\citenamefont {Alicki}\ and\ \citenamefont {Fannes}(2013)}]{Alicki13}%
  \BibitemOpen
  \bibfield  {author} {\bibinfo {author} {\bibfnamefont {R.}~\bibnamefont {Alicki}}\ and\ \bibinfo {author} {\bibfnamefont {M.}~\bibnamefont {Fannes}},\ }\bibfield  {title} {\bibinfo {title} {Entanglement boost for extractable work from ensembles of quantum batteries},\ }\href {https://doi.org/10.1103/PhysRevE.87.042123} {\bibfield  {journal} {\bibinfo  {journal} {Phys. Rev. E}\ }\textbf {\bibinfo {volume} {87}},\ \bibinfo {pages} {042123} (\bibinfo {year} {2013})}\BibitemShut {NoStop}%
\bibitem [{\citenamefont {Binder}\ \emph {et~al.}(2015)\citenamefont {Binder}, \citenamefont {Vinjanampathy}, \citenamefont {Modi},\ and\ \citenamefont {Goold}}]{Binder15}%
  \BibitemOpen
  \bibfield  {author} {\bibinfo {author} {\bibfnamefont {F.~C.}\ \bibnamefont {Binder}}, \bibinfo {author} {\bibfnamefont {S.}~\bibnamefont {Vinjanampathy}}, \bibinfo {author} {\bibfnamefont {K.}~\bibnamefont {Modi}},\ and\ \bibinfo {author} {\bibfnamefont {J.}~\bibnamefont {Goold}},\ }\bibfield  {title} {\bibinfo {title} {Quantacell: powerful charging of quantum batteries},\ }\href {https://doi.org/10.1088/1367-2630/17/7/075015} {\bibfield  {journal} {\bibinfo  {journal} {New Journal of Physics}\ }\textbf {\bibinfo {volume} {17}},\ \bibinfo {pages} {075015} (\bibinfo {year} {2015})}\BibitemShut {NoStop}%
\bibitem [{\citenamefont {Bhattacharjee}\ and\ \citenamefont {Dutta}(2021)}]{Bhattacharjee21}%
  \BibitemOpen
  \bibfield  {author} {\bibinfo {author} {\bibfnamefont {S.}~\bibnamefont {Bhattacharjee}}\ and\ \bibinfo {author} {\bibfnamefont {A.}~\bibnamefont {Dutta}},\ }\bibfield  {title} {\bibinfo {title} {Quantum thermal machines and batteries},\ }\href {https://doi.org/10.1140/epjb/s10051-021-00235-3} {\bibfield  {journal} {\bibinfo  {journal} {The European Physical Journal B}\ }\textbf {\bibinfo {volume} {94}},\ \bibinfo {pages} {239} (\bibinfo {year} {2021})}\BibitemShut {NoStop}%
\bibitem [{\citenamefont {Quach}\ \emph {et~al.}(2023)\citenamefont {Quach}, \citenamefont {Cerullo},\ and\ \citenamefont {Virgili}}]{Quach23}%
  \BibitemOpen
  \bibfield  {author} {\bibinfo {author} {\bibfnamefont {J.}~\bibnamefont {Quach}}, \bibinfo {author} {\bibfnamefont {G.}~\bibnamefont {Cerullo}},\ and\ \bibinfo {author} {\bibfnamefont {T.}~\bibnamefont {Virgili}},\ }\bibfield  {title} {\bibinfo {title} {Quantum batteries: The future of energy storage?},\ }\href {https://doi.org/https://doi.org/10.1016/j.joule.2023.09.003} {\bibfield  {journal} {\bibinfo  {journal} {Joule}\ }\textbf {\bibinfo {volume} {7}},\ \bibinfo {pages} {2195} (\bibinfo {year} {2023})}\BibitemShut {NoStop}%
\bibitem [{\citenamefont {Campaioli}\ \emph {et~al.}(2024)\citenamefont {Campaioli}, \citenamefont {Gherardini}, \citenamefont {Quach}, \citenamefont {Polini},\ and\ \citenamefont {Andolina}}]{Campaioli24}%
  \BibitemOpen
  \bibfield  {author} {\bibinfo {author} {\bibfnamefont {F.}~\bibnamefont {Campaioli}}, \bibinfo {author} {\bibfnamefont {S.}~\bibnamefont {Gherardini}}, \bibinfo {author} {\bibfnamefont {J.~Q.}\ \bibnamefont {Quach}}, \bibinfo {author} {\bibfnamefont {M.}~\bibnamefont {Polini}},\ and\ \bibinfo {author} {\bibfnamefont {G.~M.}\ \bibnamefont {Andolina}},\ }\bibfield  {title} {\bibinfo {title} {Colloquium: Quantum batteries},\ }\href {https://doi.org/10.1103/RevModPhys.96.031001} {\bibfield  {journal} {\bibinfo  {journal} {Rev. Mod. Phys.}\ }\textbf {\bibinfo {volume} {{\bf 96}}},\ \bibinfo {pages} {031001} (\bibinfo {year} {2024})}\BibitemShut {NoStop}%
\bibitem [{\citenamefont {Ferraro}\ \emph {et~al.}(2018)\citenamefont {Ferraro}, \citenamefont {Campisi}, \citenamefont {Andolina}, \citenamefont {Pellegrini},\ and\ \citenamefont {Polini}}]{Ferraro18}%
  \BibitemOpen
  \bibfield  {author} {\bibinfo {author} {\bibfnamefont {D.}~\bibnamefont {Ferraro}}, \bibinfo {author} {\bibfnamefont {M.}~\bibnamefont {Campisi}}, \bibinfo {author} {\bibfnamefont {G.~M.}\ \bibnamefont {Andolina}}, \bibinfo {author} {\bibfnamefont {V.}~\bibnamefont {Pellegrini}},\ and\ \bibinfo {author} {\bibfnamefont {M.}~\bibnamefont {Polini}},\ }\bibfield  {title} {\bibinfo {title} {High-power collective charging of a solid-state quantum battery},\ }\href@noop {} {\bibfield  {journal} {\bibinfo  {journal} {Phys. Rev. Lett.}\ }\textbf {\bibinfo {volume} {{\bf 120}}},\ \bibinfo {pages} {117702} (\bibinfo {year} {2018})}\BibitemShut {NoStop}%
\bibitem [{\citenamefont {Crescente}\ \emph {et~al.}(2020)\citenamefont {Crescente}, \citenamefont {Carrega}, \citenamefont {Sassetti},\ and\ \citenamefont {Ferraro}}]{Crescente20}%
  \BibitemOpen
  \bibfield  {author} {\bibinfo {author} {\bibfnamefont {A.}~\bibnamefont {Crescente}}, \bibinfo {author} {\bibfnamefont {M.}~\bibnamefont {Carrega}}, \bibinfo {author} {\bibfnamefont {M.}~\bibnamefont {Sassetti}},\ and\ \bibinfo {author} {\bibfnamefont {D.}~\bibnamefont {Ferraro}},\ }\bibfield  {title} {\bibinfo {title} {Ultrafast charging in a two-photon dicke quantum battery},\ }\href {https://doi.org/10.1103/PhysRevB.102.245407} {\bibfield  {journal} {\bibinfo  {journal} {Phys. Rev. B}\ }\textbf {\bibinfo {volume} {102}},\ \bibinfo {pages} {245407} (\bibinfo {year} {2020})}\BibitemShut {NoStop}%
\bibitem [{\citenamefont {Quach}\ \emph {et~al.}(2022)\citenamefont {Quach}, \citenamefont {McGhee}, \citenamefont {Ganzer}, \citenamefont {Rouse}, \citenamefont {Lovett}, \citenamefont {Gauger}, \citenamefont {Keeling}, \citenamefont {Cerullo}, \citenamefont {Lidzey},\ and\ \citenamefont {Virgili}}]{Quach22}%
  \BibitemOpen
  \bibfield  {author} {\bibinfo {author} {\bibfnamefont {J.}~\bibnamefont {Quach}}, \bibinfo {author} {\bibfnamefont {K.~E.}\ \bibnamefont {McGhee}}, \bibinfo {author} {\bibfnamefont {L.}~\bibnamefont {Ganzer}}, \bibinfo {author} {\bibfnamefont {D.~M.}\ \bibnamefont {Rouse}}, \bibinfo {author} {\bibfnamefont {B.~W.}\ \bibnamefont {Lovett}}, \bibinfo {author} {\bibfnamefont {E.~M.}\ \bibnamefont {Gauger}}, \bibinfo {author} {\bibfnamefont {J.}~\bibnamefont {Keeling}}, \bibinfo {author} {\bibfnamefont {G.}~\bibnamefont {Cerullo}}, \bibinfo {author} {\bibfnamefont {D.}~\bibnamefont {Lidzey}},\ and\ \bibinfo {author} {\bibfnamefont {T.}~\bibnamefont {Virgili}},\ }\bibfield  {title} {\bibinfo {title} {Superabsorption in an organic microcavity: Toward a quantum battery},\ }\href@noop {} {\bibfield  {journal} {\bibinfo  {journal} {Sci. Adv.}\ }\textbf {\bibinfo {volume} {{\bf 8}}},\ \bibinfo {pages} {eabk3160} (\bibinfo {year} {2022})}\BibitemShut {NoStop}%
\bibitem [{\citenamefont {Shaghaghi}\ \emph {et~al.}(2022)\citenamefont {Shaghaghi}, \citenamefont {Singh}, \citenamefont {Benenti},\ and\ \citenamefont {Rosa}}]{Shaghaghi22}%
  \BibitemOpen
  \bibfield  {author} {\bibinfo {author} {\bibfnamefont {V.}~\bibnamefont {Shaghaghi}}, \bibinfo {author} {\bibfnamefont {V.}~\bibnamefont {Singh}}, \bibinfo {author} {\bibfnamefont {G.}~\bibnamefont {Benenti}},\ and\ \bibinfo {author} {\bibfnamefont {D.}~\bibnamefont {Rosa}},\ }\bibfield  {title} {\bibinfo {title} {Micromasers as quantum batteries},\ }\href@noop {} {\bibfield  {journal} {\bibinfo  {journal} {Quantum Sci. Technol.}\ }\textbf {\bibinfo {volume} {{\bf 7}}},\ \bibinfo {pages} {04LT01} (\bibinfo {year} {2022})}\BibitemShut {NoStop}%
\bibitem [{\citenamefont {Carrasco}\ \emph {et~al.}(2022)\citenamefont {Carrasco}, \citenamefont {Maze}, \citenamefont {Hermann-Avigliano},\ and\ \citenamefont {Barra}}]{Carrasco22}%
  \BibitemOpen
  \bibfield  {author} {\bibinfo {author} {\bibfnamefont {J.}~\bibnamefont {Carrasco}}, \bibinfo {author} {\bibfnamefont {J.~R.}\ \bibnamefont {Maze}}, \bibinfo {author} {\bibfnamefont {C.}~\bibnamefont {Hermann-Avigliano}},\ and\ \bibinfo {author} {\bibfnamefont {F.}~\bibnamefont {Barra}},\ }\bibfield  {title} {\bibinfo {title} {Collective enhancement in dissipative quantum batteries},\ }\href {https://doi.org/10.1103/PhysRevE.105.064119} {\bibfield  {journal} {\bibinfo  {journal} {Phys. Rev. E}\ }\textbf {\bibinfo {volume} {105}},\ \bibinfo {pages} {064119} (\bibinfo {year} {2022})}\BibitemShut {NoStop}%
\bibitem [{\citenamefont {Gemme}\ \emph {et~al.}(2023)\citenamefont {Gemme}, \citenamefont {Andolina}, \citenamefont {Pellegrino}, \citenamefont {Sassetti},\ and\ \citenamefont {Ferraro}}]{Gemme23}%
  \BibitemOpen
  \bibfield  {author} {\bibinfo {author} {\bibfnamefont {G.}~\bibnamefont {Gemme}}, \bibinfo {author} {\bibfnamefont {G.~M.}\ \bibnamefont {Andolina}}, \bibinfo {author} {\bibfnamefont {F.~M.~D.}\ \bibnamefont {Pellegrino}}, \bibinfo {author} {\bibfnamefont {M.}~\bibnamefont {Sassetti}},\ and\ \bibinfo {author} {\bibfnamefont {D.}~\bibnamefont {Ferraro}},\ }\bibfield  {title} {\bibinfo {title} {Off-resonant dicke quantum battery: Charging by virtual photons},\ }\href {https://www.mdpi.com/2313-0105/9/4/197} {\bibfield  {journal} {\bibinfo  {journal} {Batteries}\ }\textbf {\bibinfo {volume} {9}} (\bibinfo {year} {2023})}\BibitemShut {NoStop}%
\bibitem [{\citenamefont {Rodr\'{\i}guez}\ \emph {et~al.}(2023)\citenamefont {Rodr\'{\i}guez}, \citenamefont {Rosa},\ and\ \citenamefont {Olle}}]{Rodriguez23}%
  \BibitemOpen
  \bibfield  {author} {\bibinfo {author} {\bibfnamefont {C.}~\bibnamefont {Rodr\'{\i}guez}}, \bibinfo {author} {\bibfnamefont {D.}~\bibnamefont {Rosa}},\ and\ \bibinfo {author} {\bibfnamefont {J.}~\bibnamefont {Olle}},\ }\bibfield  {title} {\bibinfo {title} {Artificial intelligence discovery of a charging protocol in a micromaser quantum battery},\ }\href {https://doi.org/10.1103/PhysRevA.108.042618} {\bibfield  {journal} {\bibinfo  {journal} {Phys. Rev. A}\ }\textbf {\bibinfo {volume} {108}},\ \bibinfo {pages} {042618} (\bibinfo {year} {2023})}\BibitemShut {NoStop}%
\bibitem [{\citenamefont {Erdman}\ \emph {et~al.}(2024)\citenamefont {Erdman}, \citenamefont {Andolina}, \citenamefont {Giovannetti},\ and\ \citenamefont {No\'e}}]{Erdman24}%
  \BibitemOpen
  \bibfield  {author} {\bibinfo {author} {\bibfnamefont {P.~A.}\ \bibnamefont {Erdman}}, \bibinfo {author} {\bibfnamefont {G.~M.}\ \bibnamefont {Andolina}}, \bibinfo {author} {\bibfnamefont {V.}~\bibnamefont {Giovannetti}},\ and\ \bibinfo {author} {\bibfnamefont {F.}~\bibnamefont {No\'e}},\ }\bibfield  {title} {\bibinfo {title} {Reinforcement learning optimization of the charging of a dicke quantum battery},\ }\href {https://doi.org/10.1103/PhysRevLett.133.243602} {\bibfield  {journal} {\bibinfo  {journal} {Phys. Rev. Lett.}\ }\textbf {\bibinfo {volume} {133}},\ \bibinfo {pages} {243602} (\bibinfo {year} {2024})}\BibitemShut {NoStop}%
\bibitem [{\citenamefont {Wang}\ \emph {et~al.}(2024)\citenamefont {Wang}, \citenamefont {Liu}, \citenamefont {Wu}, \citenamefont {Fan},\ and\ \citenamefont {Liu}}]{Wang24}%
  \BibitemOpen
  \bibfield  {author} {\bibinfo {author} {\bibfnamefont {L.}~\bibnamefont {Wang}}, \bibinfo {author} {\bibfnamefont {S.-Q.}\ \bibnamefont {Liu}}, \bibinfo {author} {\bibfnamefont {F.-l.}\ \bibnamefont {Wu}}, \bibinfo {author} {\bibfnamefont {H.}~\bibnamefont {Fan}},\ and\ \bibinfo {author} {\bibfnamefont {S.-Y.}\ \bibnamefont {Liu}},\ }\bibfield  {title} {\bibinfo {title} {Deep strong charging in a multiphoton anisotropic dicke quantum battery},\ }\href {https://doi.org/10.1103/PhysRevA.110.042419} {\bibfield  {journal} {\bibinfo  {journal} {Phys. Rev. A}\ }\textbf {\bibinfo {volume} {110}},\ \bibinfo {pages} {042419} (\bibinfo {year} {2024})}\BibitemShut {NoStop}%
\bibitem [{\citenamefont {Canzio}\ \emph {et~al.}(2025)\citenamefont {Canzio}, \citenamefont {Cavina}, \citenamefont {Polini},\ and\ \citenamefont {Giovannetti}}]{Canzio25}%
  \BibitemOpen
  \bibfield  {author} {\bibinfo {author} {\bibfnamefont {A.}~\bibnamefont {Canzio}}, \bibinfo {author} {\bibfnamefont {V.}~\bibnamefont {Cavina}}, \bibinfo {author} {\bibfnamefont {M.}~\bibnamefont {Polini}},\ and\ \bibinfo {author} {\bibfnamefont {V.}~\bibnamefont {Giovannetti}},\ }\bibfield  {title} {\bibinfo {title} {Single-atom dissipation and dephasing in dicke and tavis-cummings quantum batteries},\ }\href {https://doi.org/10.1103/PhysRevA.111.022222} {\bibfield  {journal} {\bibinfo  {journal} {Phys. Rev. A}\ }\textbf {\bibinfo {volume} {111}},\ \bibinfo {pages} {022222} (\bibinfo {year} {2025})}\BibitemShut {NoStop}%
\bibitem [{\citenamefont {Massa}\ \emph {et~al.}(2025)\citenamefont {Massa}, \citenamefont {Cavaliere},\ and\ \citenamefont {Ferraro}}]{Massa25}%
  \BibitemOpen
  \bibfield  {author} {\bibinfo {author} {\bibfnamefont {N.}~\bibnamefont {Massa}}, \bibinfo {author} {\bibfnamefont {F.}~\bibnamefont {Cavaliere}},\ and\ \bibinfo {author} {\bibfnamefont {D.}~\bibnamefont {Ferraro}},\ }\bibfield  {title} {\bibinfo {title} {The collisional charging of a transmon quantum battery},\ }\href@noop {} {\bibfield  {journal} {\bibinfo  {journal} {Batteries}\ }\textbf {\bibinfo {volume} {{\bf 11}}},\ \bibinfo {pages} {240} (\bibinfo {year} {2025})}\BibitemShut {NoStop}%
\bibitem [{\citenamefont {Hymas}\ \emph {et~al.}(2025)\citenamefont {Hymas}, \citenamefont {Muir}, \citenamefont {Tibben}, \citenamefont {van Embden}, \citenamefont {Hirai}, \citenamefont {Dunn}, \citenamefont {Gómez}, \citenamefont {Hutchison}, \citenamefont {Smith},\ and\ \citenamefont {Quach}}]{Hymas25}%
  \BibitemOpen
  \bibfield  {author} {\bibinfo {author} {\bibfnamefont {K.}~\bibnamefont {Hymas}}, \bibinfo {author} {\bibfnamefont {J.~B.}\ \bibnamefont {Muir}}, \bibinfo {author} {\bibfnamefont {D.}~\bibnamefont {Tibben}}, \bibinfo {author} {\bibfnamefont {J.}~\bibnamefont {van Embden}}, \bibinfo {author} {\bibfnamefont {T.}~\bibnamefont {Hirai}}, \bibinfo {author} {\bibfnamefont {C.~J.}\ \bibnamefont {Dunn}}, \bibinfo {author} {\bibfnamefont {D.~E.}\ \bibnamefont {Gómez}}, \bibinfo {author} {\bibfnamefont {J.~A.}\ \bibnamefont {Hutchison}}, \bibinfo {author} {\bibfnamefont {T.~A.}\ \bibnamefont {Smith}},\ and\ \bibinfo {author} {\bibfnamefont {J.~Q.}\ \bibnamefont {Quach}},\ }\href {https://arxiv.org/abs/2501.16541} {\bibinfo {title} {Experimental demonstration of a scalable room-temperature quantum battery}} (\bibinfo {year} {2025}),\ \Eprint {https://arxiv.org/abs/2501.16541} {arXiv:2501.16541 [quant-ph]} \BibitemShut {NoStop}%
\bibitem [{\citenamefont {Kurman}\ \emph {et~al.}(2025)\citenamefont {Kurman}, \citenamefont {Hymas}, \citenamefont {Fedorov}, \citenamefont {Munro},\ and\ \citenamefont {Quach}}]{Kurman25}%
  \BibitemOpen
  \bibfield  {author} {\bibinfo {author} {\bibfnamefont {Y.}~\bibnamefont {Kurman}}, \bibinfo {author} {\bibfnamefont {K.}~\bibnamefont {Hymas}}, \bibinfo {author} {\bibfnamefont {A.}~\bibnamefont {Fedorov}}, \bibinfo {author} {\bibfnamefont {W.~J.}\ \bibnamefont {Munro}},\ and\ \bibinfo {author} {\bibfnamefont {J.}~\bibnamefont {Quach}},\ }\href {https://arxiv.org/abs/2503.23610} {\bibinfo {title} {Quantum computation with quantum batteries}} (\bibinfo {year} {2025}),\ \Eprint {https://arxiv.org/abs/2503.23610} {arXiv:2503.23610} \BibitemShut {NoStop}%
\bibitem [{\citenamefont {Le}\ \emph {et~al.}(2018)\citenamefont {Le}, \citenamefont {Levinsen}, \citenamefont {Modi}, \citenamefont {Parish},\ and\ \citenamefont {Pollock}}]{Le18}%
  \BibitemOpen
  \bibfield  {author} {\bibinfo {author} {\bibfnamefont {T.~P.}\ \bibnamefont {Le}}, \bibinfo {author} {\bibfnamefont {J.}~\bibnamefont {Levinsen}}, \bibinfo {author} {\bibfnamefont {K.}~\bibnamefont {Modi}}, \bibinfo {author} {\bibfnamefont {M.~M.}\ \bibnamefont {Parish}},\ and\ \bibinfo {author} {\bibfnamefont {F.~A.}\ \bibnamefont {Pollock}},\ }\bibfield  {title} {\bibinfo {title} {Spin-chain model of a many-body quantum battery},\ }\href {https://doi.org/10.1103/PhysRevA.97.022106} {\bibfield  {journal} {\bibinfo  {journal} {Phys. Rev. A}\ }\textbf {\bibinfo {volume} {{\bf 97}}},\ \bibinfo {pages} {022106} (\bibinfo {year} {2018})}\BibitemShut {NoStop}%
\bibitem [{\citenamefont {Rossini}\ \emph {et~al.}(2020)\citenamefont {Rossini}, \citenamefont {Andolina}, \citenamefont {Rosa}, \citenamefont {Carrega},\ and\ \citenamefont {Polini}}]{Rossini20}%
  \BibitemOpen
  \bibfield  {author} {\bibinfo {author} {\bibfnamefont {D.}~\bibnamefont {Rossini}}, \bibinfo {author} {\bibfnamefont {G.~M.}\ \bibnamefont {Andolina}}, \bibinfo {author} {\bibfnamefont {D.}~\bibnamefont {Rosa}}, \bibinfo {author} {\bibfnamefont {M.}~\bibnamefont {Carrega}},\ and\ \bibinfo {author} {\bibfnamefont {M.}~\bibnamefont {Polini}},\ }\bibfield  {title} {\bibinfo {title} {Quantum advantage in the charging process of sachdev-ye-kitaev batteries},\ }\href {https://doi.org/10.1103/PhysRevLett.125.236402} {\bibfield  {journal} {\bibinfo  {journal} {Phys. Rev. Lett.}\ }\textbf {\bibinfo {volume} {125}},\ \bibinfo {pages} {236402} (\bibinfo {year} {2020})}\BibitemShut {NoStop}%
\bibitem [{\citenamefont {Catalano}\ \emph {et~al.}(2024)\citenamefont {Catalano}, \citenamefont {Giampaolo}, \citenamefont {Morsch}, \citenamefont {Giovannetti},\ and\ \citenamefont {Franchini}}]{Catalano24}%
  \BibitemOpen
  \bibfield  {author} {\bibinfo {author} {\bibfnamefont {A.}~\bibnamefont {Catalano}}, \bibinfo {author} {\bibfnamefont {S.}~\bibnamefont {Giampaolo}}, \bibinfo {author} {\bibfnamefont {O.}~\bibnamefont {Morsch}}, \bibinfo {author} {\bibfnamefont {V.}~\bibnamefont {Giovannetti}},\ and\ \bibinfo {author} {\bibfnamefont {F.}~\bibnamefont {Franchini}},\ }\bibfield  {title} {\bibinfo {title} {Frustrating quantum batteries},\ }\href {https://doi.org/10.1103/PRXQuantum.5.030319} {\bibfield  {journal} {\bibinfo  {journal} {PRX Quantum}\ }\textbf {\bibinfo {volume} {{\bf 5}}},\ \bibinfo {pages} {030319} (\bibinfo {year} {2024})}\BibitemShut {NoStop}%
\bibitem [{\citenamefont {Grazi}\ \emph {et~al.}(2024)\citenamefont {Grazi}, \citenamefont {Sacco~Shaikh}, \citenamefont {Sassetti}, \citenamefont {Traverso~Ziani},\ and\ \citenamefont {Ferraro}}]{Grazi24}%
  \BibitemOpen
  \bibfield  {author} {\bibinfo {author} {\bibfnamefont {R.}~\bibnamefont {Grazi}}, \bibinfo {author} {\bibfnamefont {D.}~\bibnamefont {Sacco~Shaikh}}, \bibinfo {author} {\bibfnamefont {M.}~\bibnamefont {Sassetti}}, \bibinfo {author} {\bibfnamefont {N.}~\bibnamefont {Traverso~Ziani}},\ and\ \bibinfo {author} {\bibfnamefont {D.}~\bibnamefont {Ferraro}},\ }\bibfield  {title} {\bibinfo {title} {Controlling energy storage crossing quantum phase transitions in an integrable spin quantum battery},\ }\href@noop {} {\bibfield  {journal} {\bibinfo  {journal} {Phys. Rev. Lett.}\ }\textbf {\bibinfo {volume} {{\bf 133}}},\ \bibinfo {pages} {197001} (\bibinfo {year} {2024})}\BibitemShut {NoStop}%
\bibitem [{\citenamefont {Grazi}\ \emph {et~al.}(2025)\citenamefont {Grazi}, \citenamefont {Cavaliere}, \citenamefont {Sassetti}, \citenamefont {Ferraro},\ and\ \citenamefont {{Traverso Ziani}}}]{Grazi25}%
  \BibitemOpen
  \bibfield  {author} {\bibinfo {author} {\bibfnamefont {R.}~\bibnamefont {Grazi}}, \bibinfo {author} {\bibfnamefont {F.}~\bibnamefont {Cavaliere}}, \bibinfo {author} {\bibfnamefont {M.}~\bibnamefont {Sassetti}}, \bibinfo {author} {\bibfnamefont {D.}~\bibnamefont {Ferraro}},\ and\ \bibinfo {author} {\bibfnamefont {N.}~\bibnamefont {{Traverso Ziani}}},\ }\bibfield  {title} {\bibinfo {title} {Charging free fermion quantum batteries},\ }\href {https://doi.org/https://doi.org/10.1016/j.chaos.2025.116383} {\bibfield  {journal} {\bibinfo  {journal} {Chaos, Solitons \& Fractals}\ }\textbf {\bibinfo {volume} {196}},\ \bibinfo {pages} {116383} (\bibinfo {year} {2025})}\BibitemShut {NoStop}%
\bibitem [{\citenamefont {Donelli}\ \emph {et~al.}(2025)\citenamefont {Donelli}, \citenamefont {Gherardini}, \citenamefont {Marino}, \citenamefont {Campaioli},\ and\ \citenamefont {Buffoni}}]{Donelli25}%
  \BibitemOpen
  \bibfield  {author} {\bibinfo {author} {\bibfnamefont {B.}~\bibnamefont {Donelli}}, \bibinfo {author} {\bibfnamefont {S.}~\bibnamefont {Gherardini}}, \bibinfo {author} {\bibfnamefont {R.}~\bibnamefont {Marino}}, \bibinfo {author} {\bibfnamefont {F.}~\bibnamefont {Campaioli}},\ and\ \bibinfo {author} {\bibfnamefont {L.}~\bibnamefont {Buffoni}},\ }\bibfield  {title} {\bibinfo {title} {Charging a quantum spin network with superextensive precision},\ }\href {https://doi.org/10.1103/lffq-ylgz} {\bibfield  {journal} {\bibinfo  {journal} {Phys. Rev. E}\ }\textbf {\bibinfo {volume} {111}},\ \bibinfo {pages} {L062102} (\bibinfo {year} {2025})}\BibitemShut {NoStop}%
\bibitem [{\citenamefont {Beder}\ \emph {et~al.}(2025)\citenamefont {Beder}, \citenamefont {Ferraro},\ and\ \citenamefont {Brandão}}]{Beder25}%
  \BibitemOpen
  \bibfield  {author} {\bibinfo {author} {\bibfnamefont {I.}~\bibnamefont {Beder}}, \bibinfo {author} {\bibfnamefont {D.}~\bibnamefont {Ferraro}},\ and\ \bibinfo {author} {\bibfnamefont {P.~A.}\ \bibnamefont {Brandão}},\ }\href {https://arxiv.org/abs/2508.19135} {\bibinfo {title} {Work extraction from a quantum battery charged through an array of coupled cavities}} (\bibinfo {year} {2025}),\ \Eprint {https://arxiv.org/abs/2508.19135} {arXiv:2508.19135 [quant-ph]} \BibitemShut {NoStop}%
\bibitem [{\citenamefont {Joshi}\ and\ \citenamefont {Mahesh}(2022)}]{Joshi22}%
  \BibitemOpen
  \bibfield  {author} {\bibinfo {author} {\bibfnamefont {J.}~\bibnamefont {Joshi}}\ and\ \bibinfo {author} {\bibfnamefont {T.~S.}\ \bibnamefont {Mahesh}},\ }\bibfield  {title} {\bibinfo {title} {Experimental investigation of a quantum battery using star-topology nmr spin systems},\ }\href {https://doi.org/10.1103/PhysRevA.106.042601} {\bibfield  {journal} {\bibinfo  {journal} {Phys. Rev. A}\ }\textbf {\bibinfo {volume} {106}},\ \bibinfo {pages} {042601} (\bibinfo {year} {2022})}\BibitemShut {NoStop}%
\bibitem [{\citenamefont {Cruz}\ \emph {et~al.}(2022)\citenamefont {Cruz}, \citenamefont {Anka}, \citenamefont {Reis}, \citenamefont {Bachelard},\ and\ \citenamefont {Santos}}]{Cruz22}%
  \BibitemOpen
  \bibfield  {author} {\bibinfo {author} {\bibfnamefont {C.}~\bibnamefont {Cruz}}, \bibinfo {author} {\bibfnamefont {M.~F.}\ \bibnamefont {Anka}}, \bibinfo {author} {\bibfnamefont {M.~S.}\ \bibnamefont {Reis}}, \bibinfo {author} {\bibfnamefont {R.}~\bibnamefont {Bachelard}},\ and\ \bibinfo {author} {\bibfnamefont {A.~C.}\ \bibnamefont {Santos}},\ }\bibfield  {title} {\bibinfo {title} {Quantum battery based on quantum discord at room temperature},\ }\href {https://doi.org/10.1088/2058-9565/ac57f3} {\bibfield  {journal} {\bibinfo  {journal} {Quantum Science and Technology}\ }\textbf {\bibinfo {volume} {7}},\ \bibinfo {pages} {025020} (\bibinfo {year} {2022})}\BibitemShut {NoStop}%
\bibitem [{\citenamefont {Andolina}\ \emph {et~al.}(2018)\citenamefont {Andolina}, \citenamefont {Farina}, \citenamefont {Mari}, \citenamefont {Pellegrini}, \citenamefont {Giovannetti},\ and\ \citenamefont {Polini}}]{Andolina18}%
  \BibitemOpen
  \bibfield  {author} {\bibinfo {author} {\bibfnamefont {G.~M.}\ \bibnamefont {Andolina}}, \bibinfo {author} {\bibfnamefont {D.}~\bibnamefont {Farina}}, \bibinfo {author} {\bibfnamefont {A.}~\bibnamefont {Mari}}, \bibinfo {author} {\bibfnamefont {V.}~\bibnamefont {Pellegrini}}, \bibinfo {author} {\bibfnamefont {V.}~\bibnamefont {Giovannetti}},\ and\ \bibinfo {author} {\bibfnamefont {M.}~\bibnamefont {Polini}},\ }\bibfield  {title} {\bibinfo {title} {Charger-mediated energy transfer in exactly solvable models for quantum batteries},\ }\href@noop {} {\bibfield  {journal} {\bibinfo  {journal} {Phys. Rev. B}\ }\textbf {\bibinfo {volume} {{\bf 98}}},\ \bibinfo {pages} {205423} (\bibinfo {year} {2018})}\BibitemShut {NoStop}%
\bibitem [{\citenamefont {Hovhannisyan}\ \emph {et~al.}(2020{\natexlab{a}})\citenamefont {Hovhannisyan}, \citenamefont {Barra},\ and\ \citenamefont {Imparato}}]{Barra20}%
  \BibitemOpen
  \bibfield  {author} {\bibinfo {author} {\bibfnamefont {K.~V.}\ \bibnamefont {Hovhannisyan}}, \bibinfo {author} {\bibfnamefont {F.}~\bibnamefont {Barra}},\ and\ \bibinfo {author} {\bibfnamefont {A.}~\bibnamefont {Imparato}},\ }\bibfield  {title} {\bibinfo {title} {Charging assisted by thermalization},\ }\href@noop {} {\bibfield  {journal} {\bibinfo  {journal} {Phys. Rev. Res.}\ }\textbf {\bibinfo {volume} {{\bf 2}}},\ \bibinfo {pages} {033413} (\bibinfo {year} {2020}{\natexlab{a}})}\BibitemShut {NoStop}%
\bibitem [{\citenamefont {Downing}\ and\ \citenamefont {Ukhtary}(2023)}]{Downing23}%
  \BibitemOpen
  \bibfield  {author} {\bibinfo {author} {\bibfnamefont {C.~A.}\ \bibnamefont {Downing}}\ and\ \bibinfo {author} {\bibfnamefont {M.~S.}\ \bibnamefont {Ukhtary}},\ }\bibfield  {title} {\bibinfo {title} {A quantum battery with quadratic driving},\ }\href {https://doi.org/10.1038/s42005-023-01439-y} {\bibfield  {journal} {\bibinfo  {journal} {Communications Physics}\ }\textbf {\bibinfo {volume} {6}},\ \bibinfo {pages} {322} (\bibinfo {year} {2023})}\BibitemShut {NoStop}%
\bibitem [{\citenamefont {Konar}\ \emph {et~al.}(2024)\citenamefont {Konar}, \citenamefont {Patra}, \citenamefont {Gupta}, \citenamefont {Ghosh},\ and\ \citenamefont {Sen(De)}}]{Konar24}%
  \BibitemOpen
  \bibfield  {author} {\bibinfo {author} {\bibfnamefont {T.~K.}\ \bibnamefont {Konar}}, \bibinfo {author} {\bibfnamefont {A.}~\bibnamefont {Patra}}, \bibinfo {author} {\bibfnamefont {R.}~\bibnamefont {Gupta}}, \bibinfo {author} {\bibfnamefont {S.}~\bibnamefont {Ghosh}},\ and\ \bibinfo {author} {\bibfnamefont {A.}~\bibnamefont {Sen(De)}},\ }\bibfield  {title} {\bibinfo {title} {Multimode advantage in continuous-variable quantum batteries},\ }\href {https://doi.org/10.1103/PhysRevA.110.022226} {\bibfield  {journal} {\bibinfo  {journal} {Phys. Rev. A}\ }\textbf {\bibinfo {volume} {110}},\ \bibinfo {pages} {022226} (\bibinfo {year} {2024})}\BibitemShut {NoStop}%
\bibitem [{\citenamefont {Cavaliere}\ \emph {et~al.}(2025)\citenamefont {Cavaliere}, \citenamefont {Gemme}, \citenamefont {Benenti}, \citenamefont {Ferraro},\ and\ \citenamefont {Sassetti}}]{Cavaliere25}%
  \BibitemOpen
  \bibfield  {author} {\bibinfo {author} {\bibfnamefont {F.}~\bibnamefont {Cavaliere}}, \bibinfo {author} {\bibfnamefont {G.}~\bibnamefont {Gemme}}, \bibinfo {author} {\bibfnamefont {G.}~\bibnamefont {Benenti}}, \bibinfo {author} {\bibfnamefont {D.}~\bibnamefont {Ferraro}},\ and\ \bibinfo {author} {\bibfnamefont {M.}~\bibnamefont {Sassetti}},\ }\bibfield  {title} {\bibinfo {title} {Dynamical blockade of a reservoir for optimal performances of a quantum battery},\ }\href@noop {} {\bibfield  {journal} {\bibinfo  {journal} {Commun. Phys.}\ }\textbf {\bibinfo {volume} {{\bf 8}}},\ \bibinfo {pages} {76} (\bibinfo {year} {2025})}\BibitemShut {NoStop}%
\bibitem [{\citenamefont {Andolina}\ \emph {et~al.}(2025)\citenamefont {Andolina}, \citenamefont {Stanzione}, \citenamefont {Giovannetti},\ and\ \citenamefont {Polini}}]{Andolina25}%
  \BibitemOpen
  \bibfield  {author} {\bibinfo {author} {\bibfnamefont {G.~M.}\ \bibnamefont {Andolina}}, \bibinfo {author} {\bibfnamefont {V.}~\bibnamefont {Stanzione}}, \bibinfo {author} {\bibfnamefont {V.}~\bibnamefont {Giovannetti}},\ and\ \bibinfo {author} {\bibfnamefont {M.}~\bibnamefont {Polini}},\ }\bibfield  {title} {\bibinfo {title} {Genuine quantum advantage in anharmonic bosonic quantum batteries},\ }\href {https://doi.org/10.1103/kzvn-dj7v} {\bibfield  {journal} {\bibinfo  {journal} {Phys. Rev. Lett.}\ }\textbf {\bibinfo {volume} {134}},\ \bibinfo {pages} {240403} (\bibinfo {year} {2025})}\BibitemShut {NoStop}%
\bibitem [{\citenamefont {Juli\`a-Farr\'e}\ \emph {et~al.}(2020)\citenamefont {Juli\`a-Farr\'e}, \citenamefont {Salamon}, \citenamefont {Riera}, \citenamefont {Bera},\ and\ \citenamefont {Lewenstein}}]{Julia20}%
  \BibitemOpen
  \bibfield  {author} {\bibinfo {author} {\bibfnamefont {S.}~\bibnamefont {Juli\`a-Farr\'e}}, \bibinfo {author} {\bibfnamefont {T.}~\bibnamefont {Salamon}}, \bibinfo {author} {\bibfnamefont {A.}~\bibnamefont {Riera}}, \bibinfo {author} {\bibfnamefont {M.~N.}\ \bibnamefont {Bera}},\ and\ \bibinfo {author} {\bibfnamefont {M.}~\bibnamefont {Lewenstein}},\ }\bibfield  {title} {\bibinfo {title} {Bounds on the capacity and power of quantum batteries},\ }\href {https://doi.org/10.1103/PhysRevResearch.2.023113} {\bibfield  {journal} {\bibinfo  {journal} {Phys. Rev. Res.}\ }\textbf {\bibinfo {volume} {2}},\ \bibinfo {pages} {023113} (\bibinfo {year} {2020})}\BibitemShut {NoStop}%
\bibitem [{\citenamefont {Andolina}\ \emph {et~al.}(2019)\citenamefont {Andolina}, \citenamefont {Keck}, \citenamefont {Mari}, \citenamefont {Giovannetti},\ and\ \citenamefont {Polini}}]{Andolina19}%
  \BibitemOpen
  \bibfield  {author} {\bibinfo {author} {\bibfnamefont {G.~M.}\ \bibnamefont {Andolina}}, \bibinfo {author} {\bibfnamefont {M.}~\bibnamefont {Keck}}, \bibinfo {author} {\bibfnamefont {A.}~\bibnamefont {Mari}}, \bibinfo {author} {\bibfnamefont {V.}~\bibnamefont {Giovannetti}},\ and\ \bibinfo {author} {\bibfnamefont {M.}~\bibnamefont {Polini}},\ }\bibfield  {title} {\bibinfo {title} {Quantum versus classical many-body batteries},\ }\href@noop {} {\bibfield  {journal} {\bibinfo  {journal} {Phys. Rev. B}\ }\textbf {\bibinfo {volume} {{\bf 99}}},\ \bibinfo {pages} {205437} (\bibinfo {year} {2019})}\BibitemShut {NoStop}%
\bibitem [{\citenamefont {Hovhannisyan}\ \emph {et~al.}(2020{\natexlab{b}})\citenamefont {Hovhannisyan}, \citenamefont {Barra},\ and\ \citenamefont {Imparato}}]{Hovhannisyan20}%
  \BibitemOpen
  \bibfield  {author} {\bibinfo {author} {\bibfnamefont {K.~V.}\ \bibnamefont {Hovhannisyan}}, \bibinfo {author} {\bibfnamefont {F.}~\bibnamefont {Barra}},\ and\ \bibinfo {author} {\bibfnamefont {A.}~\bibnamefont {Imparato}},\ }\bibfield  {title} {\bibinfo {title} {Charging assisted by thermalization},\ }\href {https://doi.org/10.1103/PhysRevResearch.2.033413} {\bibfield  {journal} {\bibinfo  {journal} {Phys. Rev. Res.}\ }\textbf {\bibinfo {volume} {2}},\ \bibinfo {pages} {033413} (\bibinfo {year} {2020}{\natexlab{b}})}\BibitemShut {NoStop}%
\bibitem [{\citenamefont {Campaioli}\ \emph {et~al.}(2017)\citenamefont {Campaioli}, \citenamefont {Pollock}, \citenamefont {Binder}, \citenamefont {C\'eleri}, \citenamefont {Goold}, \citenamefont {Vinjanampathy},\ and\ \citenamefont {Modi}}]{Campaioli17}%
  \BibitemOpen
  \bibfield  {author} {\bibinfo {author} {\bibfnamefont {F.}~\bibnamefont {Campaioli}}, \bibinfo {author} {\bibfnamefont {F.~A.}\ \bibnamefont {Pollock}}, \bibinfo {author} {\bibfnamefont {F.~C.}\ \bibnamefont {Binder}}, \bibinfo {author} {\bibfnamefont {L.}~\bibnamefont {C\'eleri}}, \bibinfo {author} {\bibfnamefont {J.}~\bibnamefont {Goold}}, \bibinfo {author} {\bibfnamefont {S.}~\bibnamefont {Vinjanampathy}},\ and\ \bibinfo {author} {\bibfnamefont {K.}~\bibnamefont {Modi}},\ }\bibfield  {title} {\bibinfo {title} {Enhancing the charging power of quantum batteries},\ }\href {https://doi.org/10.1103/PhysRevLett.118.150601} {\bibfield  {journal} {\bibinfo  {journal} {Phys. Rev. Lett.}\ }\textbf {\bibinfo {volume} {118}},\ \bibinfo {pages} {150601} (\bibinfo {year} {2017})}\BibitemShut {NoStop}%
\bibitem [{\citenamefont {Gyhm}\ \emph {et~al.}(2022)\citenamefont {Gyhm}, \citenamefont {\ifmmode~\check{S}\else \v{S}\fi{}afr\'anek},\ and\ \citenamefont {Rosa}}]{Gyhm22}%
  \BibitemOpen
  \bibfield  {author} {\bibinfo {author} {\bibfnamefont {J.-Y.}\ \bibnamefont {Gyhm}}, \bibinfo {author} {\bibfnamefont {D.}~\bibnamefont {\ifmmode~\check{S}\else \v{S}\fi{}afr\'anek}},\ and\ \bibinfo {author} {\bibfnamefont {D.}~\bibnamefont {Rosa}},\ }\bibfield  {title} {\bibinfo {title} {Quantum charging advantage cannot be extensive without global operations},\ }\href {https://doi.org/10.1103/PhysRevLett.128.140501} {\bibfield  {journal} {\bibinfo  {journal} {Phys. Rev. Lett.}\ }\textbf {\bibinfo {volume} {128}},\ \bibinfo {pages} {140501} (\bibinfo {year} {2022})}\BibitemShut {NoStop}%
\bibitem [{\citenamefont {Rinaldi}\ \emph {et~al.}(2025)\citenamefont {Rinaldi}, \citenamefont {Filip}, \citenamefont {Gerace},\ and\ \citenamefont {Guarnieri}}]{Rinaldi24}%
  \BibitemOpen
  \bibfield  {author} {\bibinfo {author} {\bibfnamefont {D.}~\bibnamefont {Rinaldi}}, \bibinfo {author} {\bibfnamefont {R.}~\bibnamefont {Filip}}, \bibinfo {author} {\bibfnamefont {D.}~\bibnamefont {Gerace}},\ and\ \bibinfo {author} {\bibfnamefont {G.}~\bibnamefont {Guarnieri}},\ }\bibfield  {title} {\bibinfo {title} {Reliable quantum advantage in quantum battery charging},\ }\href {https://doi.org/10.1103/6kwv-z6fx} {\bibfield  {journal} {\bibinfo  {journal} {Phys. Rev. A}\ }\textbf {\bibinfo {volume} {112}},\ \bibinfo {pages} {012205} (\bibinfo {year} {2025})}\BibitemShut {NoStop}%
\bibitem [{\citenamefont {Vinjanampathy}\ and\ \citenamefont {Anders}(2016)}]{Vinjanampathy2016}%
  \BibitemOpen
  \bibfield  {author} {\bibinfo {author} {\bibfnamefont {S.}~\bibnamefont {Vinjanampathy}}\ and\ \bibinfo {author} {\bibfnamefont {J.}~\bibnamefont {Anders}},\ }\bibfield  {title} {\bibinfo {title} {Quantum thermodynamics},\ }\href {https://doi.org/10.1080/00107514.2016.1201896} {\bibfield  {journal} {\bibinfo  {journal} {Contemporary Physics}\ }\textbf {\bibinfo {volume} {57}},\ \bibinfo {pages} {545} (\bibinfo {year} {2016})}\BibitemShut {NoStop}%
\bibitem [{\citenamefont {Benenti}\ \emph {et~al.}(2017)\citenamefont {Benenti}, \citenamefont {Casati}, \citenamefont {Saito},\ and\ \citenamefont {Whitney}}]{Benenti2017}%
  \BibitemOpen
  \bibfield  {author} {\bibinfo {author} {\bibfnamefont {G.}~\bibnamefont {Benenti}}, \bibinfo {author} {\bibfnamefont {G.}~\bibnamefont {Casati}}, \bibinfo {author} {\bibfnamefont {K.}~\bibnamefont {Saito}},\ and\ \bibinfo {author} {\bibfnamefont {R.}~\bibnamefont {Whitney}},\ }\bibfield  {title} {\bibinfo {title} {Fundamental aspects of steady-state conversion of heat to work at the nanoscale},\ }\href {https://doi.org/10.1016/j.physrep.2017.05.008} {\bibfield  {journal} {\bibinfo  {journal} {Physics Reports}\ }\textbf {\bibinfo {volume} {694}},\ \bibinfo {pages} {1} (\bibinfo {year} {2017})}\BibitemShut {NoStop}%
\bibitem [{\citenamefont {Deffner}\ and\ \citenamefont {Campbell}(2019)}]{Daffner2019}%
  \BibitemOpen
  \bibfield  {author} {\bibinfo {author} {\bibfnamefont {S.}~\bibnamefont {Deffner}}\ and\ \bibinfo {author} {\bibfnamefont {S.}~\bibnamefont {Campbell}},\ }\href {https://doi.org/10.1088/2053-2571/ab21c6} {\emph {\bibinfo {title} {Quantum Thermodynamics}}},\ 2053-2571\ (\bibinfo  {publisher} {Morgan \& Claypool Publishers},\ \bibinfo {year} {2019})\BibitemShut {NoStop}%
\bibitem [{\citenamefont {Arrachea}(2023)}]{Arrachea23}%
  \BibitemOpen
  \bibfield  {author} {\bibinfo {author} {\bibfnamefont {L.}~\bibnamefont {Arrachea}},\ }\bibfield  {title} {\bibinfo {title} {Energy dynamics, heat production and heat–work conversion with qubits: toward the development of quantum machines},\ }\href {https://doi.org/10.1088/1361-6633/acb06b} {\bibfield  {journal} {\bibinfo  {journal} {Reports on Progress in Physics}\ }\textbf {\bibinfo {volume} {86}},\ \bibinfo {pages} {036501} (\bibinfo {year} {2023})}\BibitemShut {NoStop}%
\bibitem [{\citenamefont {Cangemi}\ \emph {et~al.}(2024)\citenamefont {Cangemi}, \citenamefont {Bhadra},\ and\ \citenamefont {Levy}}]{Cangemi24}%
  \BibitemOpen
  \bibfield  {author} {\bibinfo {author} {\bibfnamefont {L.~M.}\ \bibnamefont {Cangemi}}, \bibinfo {author} {\bibfnamefont {C.}~\bibnamefont {Bhadra}},\ and\ \bibinfo {author} {\bibfnamefont {A.}~\bibnamefont {Levy}},\ }\bibfield  {title} {\bibinfo {title} {Quantum engines and refrigerators},\ }\href {https://doi.org/https://doi.org/10.1016/j.physrep.2024.07.001} {\bibfield  {journal} {\bibinfo  {journal} {Physics Reports}\ }\textbf {\bibinfo {volume} {1087}},\ \bibinfo {pages} {1} (\bibinfo {year} {2024})},\ \bibinfo {note} {quantum engines and refrigerators}\BibitemShut {NoStop}%
\bibitem [{\citenamefont {Razzoli}\ \emph {et~al.}(2025)\citenamefont {Razzoli}, \citenamefont {Gemme}, \citenamefont {Khomchenko}, \citenamefont {Sassetti}, \citenamefont {Ouerdane}, \citenamefont {Ferraro},\ and\ \citenamefont {Benenti}}]{Razzoli25}%
  \BibitemOpen
  \bibfield  {author} {\bibinfo {author} {\bibfnamefont {L.}~\bibnamefont {Razzoli}}, \bibinfo {author} {\bibfnamefont {G.}~\bibnamefont {Gemme}}, \bibinfo {author} {\bibfnamefont {I.}~\bibnamefont {Khomchenko}}, \bibinfo {author} {\bibfnamefont {M.}~\bibnamefont {Sassetti}}, \bibinfo {author} {\bibfnamefont {H.}~\bibnamefont {Ouerdane}}, \bibinfo {author} {\bibfnamefont {D.}~\bibnamefont {Ferraro}},\ and\ \bibinfo {author} {\bibfnamefont {G.}~\bibnamefont {Benenti}},\ }\bibfield  {title} {\bibinfo {title} {Cyclic solid-state quantum battery: thermodynamic characterization and quantum hardware simulation},\ }\href {https://doi.org/10.1088/2058-9565/ad9ed4} {\bibfield  {journal} {\bibinfo  {journal} {Quantum Science and Technology}\ }\textbf {\bibinfo {volume} {10}},\ \bibinfo {pages} {015064} (\bibinfo {year} {2025})}\BibitemShut {NoStop}%
\bibitem [{\citenamefont {Braunstein}\ and\ \citenamefont {van Loock}(2005)}]{Braunstein05}%
  \BibitemOpen
  \bibfield  {author} {\bibinfo {author} {\bibfnamefont {S.~L.}\ \bibnamefont {Braunstein}}\ and\ \bibinfo {author} {\bibfnamefont {P.}~\bibnamefont {van Loock}},\ }\bibfield  {title} {\bibinfo {title} {Quantum information with continuous variables},\ }\href {https://doi.org/10.1103/RevModPhys.77.513} {\bibfield  {journal} {\bibinfo  {journal} {Rev. Mod. Phys.}\ }\textbf {\bibinfo {volume} {77}},\ \bibinfo {pages} {513} (\bibinfo {year} {2005})}\BibitemShut {NoStop}%
\bibitem [{\citenamefont {Adesso}\ and\ \citenamefont {Illuminati}(2007)}]{Adesso07}%
  \BibitemOpen
  \bibfield  {author} {\bibinfo {author} {\bibfnamefont {G.}~\bibnamefont {Adesso}}\ and\ \bibinfo {author} {\bibfnamefont {F.}~\bibnamefont {Illuminati}},\ }\bibfield  {title} {\bibinfo {title} {Entanglement in continuous-variable systems: recent advances and current perspectives},\ }\href {https://doi.org/10.1088/1751-8113/40/28/S01} {\bibfield  {journal} {\bibinfo  {journal} {Journal of Physics A: Mathematical and Theoretical}\ }\textbf {\bibinfo {volume} {40}},\ \bibinfo {pages} {7821} (\bibinfo {year} {2007})}\BibitemShut {NoStop}%
\bibitem [{\citenamefont {Adesso}\ \emph {et~al.}(2014)\citenamefont {Adesso}, \citenamefont {Ragy},\ and\ \citenamefont {Lee}}]{Adesso14}%
  \BibitemOpen
  \bibfield  {author} {\bibinfo {author} {\bibfnamefont {G.}~\bibnamefont {Adesso}}, \bibinfo {author} {\bibfnamefont {S.}~\bibnamefont {Ragy}},\ and\ \bibinfo {author} {\bibfnamefont {A.~R.}\ \bibnamefont {Lee}},\ }\bibfield  {title} {\bibinfo {title} {Continuous variable quantum information: Gaussian states and beyond},\ }\href {https://doi.org/10.1142/S1230161214400010} {\bibfield  {journal} {\bibinfo  {journal} {Open Systems \& Information Dynamics}\ }\textbf {\bibinfo {volume} {21}},\ \bibinfo {pages} {1440001} (\bibinfo {year} {2014})},\ \Eprint {https://arxiv.org/abs/https://doi.org/10.1142/S1230161214400010} {https://doi.org/10.1142/S1230161214400010} \BibitemShut {NoStop}%
\bibitem [{\citenamefont {Ingold}(2002)}]{Ingold02}%
  \BibitemOpen
  \bibfield  {author} {\bibinfo {author} {\bibfnamefont {G.-L.}\ \bibnamefont {Ingold}},\ }\href@noop {} {\emph {\bibinfo {title} {Lecture Notes in Physics}}},\ Vol.\ \bibinfo {volume} {611}\ (\bibinfo  {publisher} {Springer},\ \bibinfo {year} {2002})\BibitemShut {NoStop}%
\bibitem [{\citenamefont {Weiss}(2012)}]{Weiss12}%
  \BibitemOpen
  \bibfield  {author} {\bibinfo {author} {\bibfnamefont {U.}~\bibnamefont {Weiss}},\ }\href@noop {} {\emph {\bibinfo {title} {Quantum Dissipative Systems}}},\ \bibinfo {edition} {4th}\ ed.\ (\bibinfo  {publisher} {WORLD SCIENTIFIC},\ \bibinfo {year} {2012})\BibitemShut {NoStop}%
\bibitem [{\citenamefont {Freitas}\ and\ \citenamefont {Paz}(2014)}]{Freitas14}%
  \BibitemOpen
  \bibfield  {author} {\bibinfo {author} {\bibfnamefont {N.}~\bibnamefont {Freitas}}\ and\ \bibinfo {author} {\bibfnamefont {J.~P.}\ \bibnamefont {Paz}},\ }\bibfield  {title} {\bibinfo {title} {Analytic solution for heat flow through a general harmonic network},\ }\href {https://doi.org/10.1103/PhysRevE.90.042128} {\bibfield  {journal} {\bibinfo  {journal} {Phys. Rev. E}\ }\textbf {\bibinfo {volume} {90}},\ \bibinfo {pages} {042128} (\bibinfo {year} {2014})}\BibitemShut {NoStop}%
\bibitem [{\citenamefont {Carrega}\ \emph {et~al.}(2016)\citenamefont {Carrega}, \citenamefont {Solinas}, \citenamefont {Sassetti},\ and\ \citenamefont {Weiss}}]{Carrega16}%
  \BibitemOpen
  \bibfield  {author} {\bibinfo {author} {\bibfnamefont {M.}~\bibnamefont {Carrega}}, \bibinfo {author} {\bibfnamefont {P.}~\bibnamefont {Solinas}}, \bibinfo {author} {\bibfnamefont {M.}~\bibnamefont {Sassetti}},\ and\ \bibinfo {author} {\bibfnamefont {U.}~\bibnamefont {Weiss}},\ }\bibfield  {title} {\bibinfo {title} {Energy exchange in driven open quantum systems at strong coupling},\ }\href {https://doi.org/10.1103/PhysRevLett.116.240403} {\bibfield  {journal} {\bibinfo  {journal} {Phys. Rev. Lett.}\ }\textbf {\bibinfo {volume} {116}},\ \bibinfo {pages} {240403} (\bibinfo {year} {2016})}\BibitemShut {NoStop}%
\bibitem [{\citenamefont {Carrega}\ \emph {et~al.}(2020)\citenamefont {Carrega}, \citenamefont {Crescente}, \citenamefont {Ferraro},\ and\ \citenamefont {Sassetti}}]{Carrega20}%
  \BibitemOpen
  \bibfield  {author} {\bibinfo {author} {\bibfnamefont {M.}~\bibnamefont {Carrega}}, \bibinfo {author} {\bibfnamefont {A.}~\bibnamefont {Crescente}}, \bibinfo {author} {\bibfnamefont {D.}~\bibnamefont {Ferraro}},\ and\ \bibinfo {author} {\bibfnamefont {M.}~\bibnamefont {Sassetti}},\ }\bibfield  {title} {\bibinfo {title} {Dissipative dynamics of an open quantum battery},\ }\href {https://doi.org/10.1088/1367-2630/abaa01} {\bibfield  {journal} {\bibinfo  {journal} {New Journal of Physics}\ }\textbf {\bibinfo {volume} {22}},\ \bibinfo {pages} {083085} (\bibinfo {year} {2020})}\BibitemShut {NoStop}%
\bibitem [{\citenamefont {Carrega}\ \emph {et~al.}(2022)\citenamefont {Carrega}, \citenamefont {Cangemi}, \citenamefont {De~Filippis}, \citenamefont {Cataudella}, \citenamefont {Benenti},\ and\ \citenamefont {Sassetti}}]{Carrega22}%
  \BibitemOpen
  \bibfield  {author} {\bibinfo {author} {\bibfnamefont {M.}~\bibnamefont {Carrega}}, \bibinfo {author} {\bibfnamefont {L.~M.}\ \bibnamefont {Cangemi}}, \bibinfo {author} {\bibfnamefont {G.}~\bibnamefont {De~Filippis}}, \bibinfo {author} {\bibfnamefont {V.}~\bibnamefont {Cataudella}}, \bibinfo {author} {\bibfnamefont {G.}~\bibnamefont {Benenti}},\ and\ \bibinfo {author} {\bibfnamefont {M.}~\bibnamefont {Sassetti}},\ }\bibfield  {title} {\bibinfo {title} {Engineering dynamical couplings for quantum thermodynamic tasks},\ }\href {https://doi.org/10.1103/PRXQuantum.3.010323} {\bibfield  {journal} {\bibinfo  {journal} {PRX Quantum}\ }\textbf {\bibinfo {volume} {{\bf 3}, 010323}},\ \bibinfo {pages} {010323} (\bibinfo {year} {2022})}\BibitemShut {NoStop}%
\bibitem [{\citenamefont {Carrega}\ \emph {et~al.}(2024)\citenamefont {Carrega}, \citenamefont {Razzoli}, \citenamefont {Erdman}, \citenamefont {Cavaliere}, \citenamefont {Benenti},\ and\ \citenamefont {Sassetti}}]{Carrega24}%
  \BibitemOpen
  \bibfield  {author} {\bibinfo {author} {\bibfnamefont {M.}~\bibnamefont {Carrega}}, \bibinfo {author} {\bibfnamefont {L.}~\bibnamefont {Razzoli}}, \bibinfo {author} {\bibfnamefont {P.~A.}\ \bibnamefont {Erdman}}, \bibinfo {author} {\bibfnamefont {F.}~\bibnamefont {Cavaliere}}, \bibinfo {author} {\bibfnamefont {G.}~\bibnamefont {Benenti}},\ and\ \bibinfo {author} {\bibfnamefont {M.}~\bibnamefont {Sassetti}},\ }\bibfield  {title} {\bibinfo {title} {Dissipation-induced collective advantage of a quantum thermal machine},\ }\bibfield  {journal} {\bibinfo  {journal} {AVS Quantum Science}\ }\textbf {\bibinfo {volume} {6}},\ \href {https://doi.org/10.1116/5.0190340} {10.1116/5.0190340} (\bibinfo {year} {2024})\BibitemShut {NoStop}%
\bibitem [{\citenamefont {Serafini}(2017)}]{Serafini2017}%
  \BibitemOpen
  \bibfield  {author} {\bibinfo {author} {\bibfnamefont {A.}~\bibnamefont {Serafini}},\ }\href@noop {} {\emph {\bibinfo {title} {Quantum Continuous Variables: A Primer of Theoretical Methods}}}\ (\bibinfo  {publisher} {CRC Press, Taylor \& Francis},\ \bibinfo {year} {2017})\BibitemShut {NoStop}%
\bibitem [{\citenamefont {Brask}(2022)}]{Brask2022}%
  \BibitemOpen
  \bibfield  {author} {\bibinfo {author} {\bibfnamefont {J.~B.}\ \bibnamefont {Brask}},\ }\bibfield  {title} {\bibinfo {title} {Gaussian states and operations – a quick reference},\ }\href@noop {} {\bibfield  {journal} {\bibinfo  {journal} {arXiv:2102.057482v2}\ } (\bibinfo {year} {2022})}\BibitemShut {NoStop}%
\bibitem [{SM()}]{SM}%
  \BibitemOpen
  \bibfield  {journal} {\bibinfo  {journal} {See Supplemental Material}\ }\href@noop {} {}\BibitemShut {NoStop}%
\bibitem [{\citenamefont {Ciuti}\ \emph {et~al.}(2005)\citenamefont {Ciuti}, \citenamefont {Bastard},\ and\ \citenamefont {Carusotto}}]{Ciuti05}%
  \BibitemOpen
  \bibfield  {author} {\bibinfo {author} {\bibfnamefont {C.}~\bibnamefont {Ciuti}}, \bibinfo {author} {\bibfnamefont {G.}~\bibnamefont {Bastard}},\ and\ \bibinfo {author} {\bibfnamefont {I.}~\bibnamefont {Carusotto}},\ }\bibfield  {title} {\bibinfo {title} {Quantum vacuum properties of the intersubband cavity polariton field},\ }\href {https://doi.org/10.1103/PhysRevB.72.115303} {\bibfield  {journal} {\bibinfo  {journal} {Phys. Rev. B}\ }\textbf {\bibinfo {volume} {72}},\ \bibinfo {pages} {115303} (\bibinfo {year} {2005})}\BibitemShut {NoStop}%
\bibitem [{\citenamefont {Allahverdyan}\ \emph {et~al.}(2004)\citenamefont {Allahverdyan}, \citenamefont {Balian},\ and\ \citenamefont {Nieuwenhuizen}}]{Allahverdyan04}%
  \BibitemOpen
  \bibfield  {author} {\bibinfo {author} {\bibfnamefont {A.~E.}\ \bibnamefont {Allahverdyan}}, \bibinfo {author} {\bibfnamefont {R.}~\bibnamefont {Balian}},\ and\ \bibinfo {author} {\bibfnamefont {T.~M.}\ \bibnamefont {Nieuwenhuizen}},\ }\bibfield  {title} {\bibinfo {title} {Maximal work extraction from finite quantum systems},\ }\href {https://doi.org/10.1209/epl/i2004-10101-2} {\bibfield  {journal} {\bibinfo  {journal} {Europhysics Letters}\ }\textbf {\bibinfo {volume} {67}},\ \bibinfo {pages} {565} (\bibinfo {year} {2004})}\BibitemShut {NoStop}%
\bibitem [{\citenamefont {Farina}\ \emph {et~al.}(2019)\citenamefont {Farina}, \citenamefont {Andolina}, \citenamefont {Mari}, \citenamefont {Polini},\ and\ \citenamefont {Giovannetti}}]{Farina19}%
  \BibitemOpen
  \bibfield  {author} {\bibinfo {author} {\bibfnamefont {D.}~\bibnamefont {Farina}}, \bibinfo {author} {\bibfnamefont {G.~M.}\ \bibnamefont {Andolina}}, \bibinfo {author} {\bibfnamefont {A.}~\bibnamefont {Mari}}, \bibinfo {author} {\bibfnamefont {M.}~\bibnamefont {Polini}},\ and\ \bibinfo {author} {\bibfnamefont {V.}~\bibnamefont {Giovannetti}},\ }\bibfield  {title} {\bibinfo {title} {Charger-mediated energy transfer for quantum batteries: An open-system approach},\ }\href {https://doi.org/10.1103/PhysRevB.99.035421} {\bibfield  {journal} {\bibinfo  {journal} {Phys. Rev. B}\ }\textbf {\bibinfo {volume} {99}},\ \bibinfo {pages} {035421} (\bibinfo {year} {2019})}\BibitemShut {NoStop}%
\bibitem [{\citenamefont {Barra}(2022)}]{Barra22}%
  \BibitemOpen
  \bibfield  {author} {\bibinfo {author} {\bibfnamefont {F.}~\bibnamefont {Barra}},\ }\bibfield  {title} {\bibinfo {title} {Efficiency fluctuations in a quantum battery charged by a repeated interaction process},\ }\bibfield  {journal} {\bibinfo  {journal} {Entropy}\ }\textbf {\bibinfo {volume} {24}},\ \href {https://doi.org/10.3390/e24060820} {10.3390/e24060820} (\bibinfo {year} {2022})\BibitemShut {NoStop}%
\bibitem [{\citenamefont {Esposito}\ \emph {et~al.}(2010)\citenamefont {Esposito}, \citenamefont {Lindenberg},\ and\ \citenamefont {Van~den Broeck}}]{Esposito10}%
  \BibitemOpen
  \bibfield  {author} {\bibinfo {author} {\bibfnamefont {M.}~\bibnamefont {Esposito}}, \bibinfo {author} {\bibfnamefont {K.}~\bibnamefont {Lindenberg}},\ and\ \bibinfo {author} {\bibfnamefont {C.}~\bibnamefont {Van~den Broeck}},\ }\bibfield  {title} {\bibinfo {title} {Entropy production as correlation between system and reservoir},\ }\href {https://doi.org/10.1088/1367-2630/12/1/013013} {\bibfield  {journal} {\bibinfo  {journal} {New Journal of Physics}\ }\textbf {\bibinfo {volume} {12}},\ \bibinfo {pages} {013013} (\bibinfo {year} {2010})}\BibitemShut {NoStop}%
\bibitem [{\citenamefont {Rao}\ and\ \citenamefont {Esposito}(2016)}]{Rao16}%
  \BibitemOpen
  \bibfield  {author} {\bibinfo {author} {\bibfnamefont {R.}~\bibnamefont {Rao}}\ and\ \bibinfo {author} {\bibfnamefont {M.}~\bibnamefont {Esposito}},\ }\bibfield  {title} {\bibinfo {title} {Nonequilibrium thermodynamics of chemical reaction networks: Wisdom from stochastic thermodynamics},\ }\href {https://doi.org/10.1103/PhysRevX.6.041064} {\bibfield  {journal} {\bibinfo  {journal} {Phys. Rev. X}\ }\textbf {\bibinfo {volume} {6}},\ \bibinfo {pages} {041064} (\bibinfo {year} {2016})}\BibitemShut {NoStop}%
\bibitem [{\citenamefont {Landi}\ and\ \citenamefont {Paternostro}(2021)}]{Landi21}%
  \BibitemOpen
  \bibfield  {author} {\bibinfo {author} {\bibfnamefont {G.~T.}\ \bibnamefont {Landi}}\ and\ \bibinfo {author} {\bibfnamefont {M.}~\bibnamefont {Paternostro}},\ }\bibfield  {title} {\bibinfo {title} {Irreversible entropy production: From classical to quantum},\ }\href {https://doi.org/10.1103/RevModPhys.93.035008} {\bibfield  {journal} {\bibinfo  {journal} {Rev. Mod. Phys.}\ }\textbf {\bibinfo {volume} {93}},\ \bibinfo {pages} {035008} (\bibinfo {year} {2021})}\BibitemShut {NoStop}%
\bibitem [{\citenamefont {Aguilar}\ and\ \citenamefont {Lutz}(2025)}]{Aguilar25}%
  \BibitemOpen
  \bibfield  {author} {\bibinfo {author} {\bibfnamefont {M.}~\bibnamefont {Aguilar}}\ and\ \bibinfo {author} {\bibfnamefont {E.}~\bibnamefont {Lutz}},\ }\bibfield  {title} {\bibinfo {title} {Correlated quantum machines beyond the standard second law},\ }\href {https://doi.org/10.1126/sciadv.adw8462} {\bibfield  {journal} {\bibinfo  {journal} {Science Advances}\ }\textbf {\bibinfo {volume} {11}},\ \bibinfo {pages} {eadw8462} (\bibinfo {year} {2025})},\ \Eprint {https://arxiv.org/abs/https://www.science.org/doi/pdf/10.1126/sciadv.adw8462} {https://www.science.org/doi/pdf/10.1126/sciadv.adw8462} \BibitemShut {NoStop}%
\bibitem [{\citenamefont {Manzano}\ \emph {et~al.}(2020)\citenamefont {Manzano}, \citenamefont {S\'anchez}, \citenamefont {Silva}, \citenamefont {Haack}, \citenamefont {Brask}, \citenamefont {Brunner},\ and\ \citenamefont {Potts}}]{Manzano20}%
  \BibitemOpen
  \bibfield  {author} {\bibinfo {author} {\bibfnamefont {G.}~\bibnamefont {Manzano}}, \bibinfo {author} {\bibfnamefont {R.}~\bibnamefont {S\'anchez}}, \bibinfo {author} {\bibfnamefont {R.}~\bibnamefont {Silva}}, \bibinfo {author} {\bibfnamefont {G.}~\bibnamefont {Haack}}, \bibinfo {author} {\bibfnamefont {J.~B.}\ \bibnamefont {Brask}}, \bibinfo {author} {\bibfnamefont {N.}~\bibnamefont {Brunner}},\ and\ \bibinfo {author} {\bibfnamefont {P.~P.}\ \bibnamefont {Potts}},\ }\bibfield  {title} {\bibinfo {title} {Hybrid thermal machines: Generalized thermodynamic resources for multitasking},\ }\href {https://doi.org/10.1103/PhysRevResearch.2.043302} {\bibfield  {journal} {\bibinfo  {journal} {Phys. Rev. Res.}\ }\textbf {\bibinfo {volume} {2}},\ \bibinfo {pages} {043302} (\bibinfo {year} {2020})}\BibitemShut {NoStop}%
\bibitem [{\citenamefont {Lu}\ \emph {et~al.}(2023)\citenamefont {Lu}, \citenamefont {Wang}, \citenamefont {Wang}, \citenamefont {Peng}, \citenamefont {Wang},\ and\ \citenamefont {Jiang}}]{Lu23}%
  \BibitemOpen
  \bibfield  {author} {\bibinfo {author} {\bibfnamefont {J.}~\bibnamefont {Lu}}, \bibinfo {author} {\bibfnamefont {Z.}~\bibnamefont {Wang}}, \bibinfo {author} {\bibfnamefont {R.}~\bibnamefont {Wang}}, \bibinfo {author} {\bibfnamefont {J.}~\bibnamefont {Peng}}, \bibinfo {author} {\bibfnamefont {C.}~\bibnamefont {Wang}},\ and\ \bibinfo {author} {\bibfnamefont {J.-H.}\ \bibnamefont {Jiang}},\ }\bibfield  {title} {\bibinfo {title} {Multitask quantum thermal machines and cooperative effects},\ }\href {https://doi.org/10.1103/PhysRevB.107.075428} {\bibfield  {journal} {\bibinfo  {journal} {Phys. Rev. B}\ }\textbf {\bibinfo {volume} {107}},\ \bibinfo {pages} {075428} (\bibinfo {year} {2023})}\BibitemShut {NoStop}%
\bibitem [{\citenamefont {L\'opez}\ \emph {et~al.}(2023)\citenamefont {L\'opez}, \citenamefont {Lim},\ and\ \citenamefont {Kim}}]{Lopez23}%
  \BibitemOpen
  \bibfield  {author} {\bibinfo {author} {\bibfnamefont {R.}~\bibnamefont {L\'opez}}, \bibinfo {author} {\bibfnamefont {J.~S.}\ \bibnamefont {Lim}},\ and\ \bibinfo {author} {\bibfnamefont {K.~W.}\ \bibnamefont {Kim}},\ }\bibfield  {title} {\bibinfo {title} {Optimal superconducting hybrid machine},\ }\href {https://doi.org/10.1103/PhysRevResearch.5.013038} {\bibfield  {journal} {\bibinfo  {journal} {Phys. Rev. Res.}\ }\textbf {\bibinfo {volume} {5}},\ \bibinfo {pages} {013038} (\bibinfo {year} {2023})}\BibitemShut {NoStop}%
\bibitem [{\citenamefont {Cavaliere}\ \emph {et~al.}(2023)\citenamefont {Cavaliere}, \citenamefont {Razzoli}, \citenamefont {Carrega}, \citenamefont {Benenti},\ and\ \citenamefont {Sassetti}}]{Cavaliere23}%
  \BibitemOpen
  \bibfield  {author} {\bibinfo {author} {\bibfnamefont {F.}~\bibnamefont {Cavaliere}}, \bibinfo {author} {\bibfnamefont {L.}~\bibnamefont {Razzoli}}, \bibinfo {author} {\bibfnamefont {M.}~\bibnamefont {Carrega}}, \bibinfo {author} {\bibfnamefont {G.}~\bibnamefont {Benenti}},\ and\ \bibinfo {author} {\bibfnamefont {M.}~\bibnamefont {Sassetti}},\ }\bibfield  {title} {\bibinfo {title} {Hybrid quantum thermal machines with dynamical couplings},\ }\href {https://doi.org/https://doi.org/10.1016/j.isci.2023.106235} {\bibfield  {journal} {\bibinfo  {journal} {iScience}\ }\textbf {\bibinfo {volume} {26}},\ \bibinfo {pages} {106235} (\bibinfo {year} {2023})}\BibitemShut {NoStop}%
\bibitem [{\citenamefont {Finocchiaro}\ \emph {et~al.}(2025)\citenamefont {Finocchiaro}, \citenamefont {Ferraro}, \citenamefont {Sassetti},\ and\ \citenamefont {Benenti}}]{Finocchiaro25}%
  \BibitemOpen
  \bibfield  {author} {\bibinfo {author} {\bibfnamefont {S.}~\bibnamefont {Finocchiaro}}, \bibinfo {author} {\bibfnamefont {D.}~\bibnamefont {Ferraro}}, \bibinfo {author} {\bibfnamefont {M.}~\bibnamefont {Sassetti}},\ and\ \bibinfo {author} {\bibfnamefont {G.}~\bibnamefont {Benenti}},\ }\bibfield  {title} {\bibinfo {title} {Hybrid interacting quantum hall thermal machine},\ }\href {https://doi.org/10.1103/PhysRevB.111.205420} {\bibfield  {journal} {\bibinfo  {journal} {Phys. Rev. B}\ }\textbf {\bibinfo {volume} {111}},\ \bibinfo {pages} {205420} (\bibinfo {year} {2025})}\BibitemShut {NoStop}%
\bibitem [{\citenamefont {Horodecki}\ \emph {et~al.}(2009)\citenamefont {Horodecki}, \citenamefont {Horodecki}, \citenamefont {Horodecki},\ and\ \citenamefont {Horodecki}}]{Horodecki09}%
  \BibitemOpen
  \bibfield  {author} {\bibinfo {author} {\bibfnamefont {R.}~\bibnamefont {Horodecki}}, \bibinfo {author} {\bibfnamefont {P.}~\bibnamefont {Horodecki}}, \bibinfo {author} {\bibfnamefont {M.}~\bibnamefont {Horodecki}},\ and\ \bibinfo {author} {\bibfnamefont {K.}~\bibnamefont {Horodecki}},\ }\bibfield  {title} {\bibinfo {title} {Quantum entanglement},\ }\href {https://doi.org/10.1103/RevModPhys.81.865} {\bibfield  {journal} {\bibinfo  {journal} {Rev. Mod. Phys.}\ }\textbf {\bibinfo {volume} {81}},\ \bibinfo {pages} {865} (\bibinfo {year} {2009})}\BibitemShut {NoStop}%
\bibitem [{\citenamefont {Vool}\ and\ \citenamefont {Devoret}(2017)}]{Vool17}%
  \BibitemOpen
  \bibfield  {author} {\bibinfo {author} {\bibfnamefont {U.}~\bibnamefont {Vool}}\ and\ \bibinfo {author} {\bibfnamefont {M.}~\bibnamefont {Devoret}},\ }\bibfield  {title} {\bibinfo {title} {Introduction to quantum electromagnetic circuits},\ }\href {https://doi.org/https://doi.org/10.1002/cta.2359} {\bibfield  {journal} {\bibinfo  {journal} {International Journal of Circuit Theory and Applications}\ }\textbf {\bibinfo {volume} {45}},\ \bibinfo {pages} {897} (\bibinfo {year} {2017})}\BibitemShut {NoStop}%
\bibitem [{\citenamefont {Blais}\ \emph {et~al.}(2021)\citenamefont {Blais}, \citenamefont {Grimsmo}, \citenamefont {Girvin},\ and\ \citenamefont {Wallraff}}]{Blais21}%
  \BibitemOpen
  \bibfield  {author} {\bibinfo {author} {\bibfnamefont {A.}~\bibnamefont {Blais}}, \bibinfo {author} {\bibfnamefont {A.~L.}\ \bibnamefont {Grimsmo}}, \bibinfo {author} {\bibfnamefont {S.~M.}\ \bibnamefont {Girvin}},\ and\ \bibinfo {author} {\bibfnamefont {A.}~\bibnamefont {Wallraff}},\ }\bibfield  {title} {\bibinfo {title} {Circuit quantum electrodynamics},\ }\href@noop {} {\bibfield  {journal} {\bibinfo  {journal} {Rev. Mod. Phys.}\ }\textbf {\bibinfo {volume} {{\bf 93}}},\ \bibinfo {pages} {025005} (\bibinfo {year} {2021})}\BibitemShut {NoStop}%
\bibitem [{\citenamefont {Cattaneo}\ and\ \citenamefont {Paraoanu}(2021)}]{Cattaneo21}%
  \BibitemOpen
  \bibfield  {author} {\bibinfo {author} {\bibfnamefont {M.}~\bibnamefont {Cattaneo}}\ and\ \bibinfo {author} {\bibfnamefont {G.~S.}\ \bibnamefont {Paraoanu}},\ }\bibfield  {title} {\bibinfo {title} {Engineering dissipation with resistive elements in circuit quantum electrodynamics},\ }\href {https://doi.org/https://doi.org/10.1002/qute.202100054} {\bibfield  {journal} {\bibinfo  {journal} {Advanced Quantum Technologies}\ }\textbf {\bibinfo {volume} {4}},\ \bibinfo {pages} {2100054} (\bibinfo {year} {2021})}\BibitemShut {NoStop}%
\bibitem [{\citenamefont {Gramich}\ \emph {et~al.}(2011)\citenamefont {Gramich}, \citenamefont {Solinas}, \citenamefont {M\"ott\"onen}, \citenamefont {Pekola},\ and\ \citenamefont {Ankerhold}}]{Gramich11}%
  \BibitemOpen
  \bibfield  {author} {\bibinfo {author} {\bibfnamefont {V.}~\bibnamefont {Gramich}}, \bibinfo {author} {\bibfnamefont {P.}~\bibnamefont {Solinas}}, \bibinfo {author} {\bibfnamefont {M.}~\bibnamefont {M\"ott\"onen}}, \bibinfo {author} {\bibfnamefont {J.~P.}\ \bibnamefont {Pekola}},\ and\ \bibinfo {author} {\bibfnamefont {J.}~\bibnamefont {Ankerhold}},\ }\bibfield  {title} {\bibinfo {title} {Measurement scheme for the lamb shift in a superconducting circuit with broadband environment},\ }\href {https://doi.org/10.1103/PhysRevA.84.052103} {\bibfield  {journal} {\bibinfo  {journal} {Phys. Rev. A}\ }\textbf {\bibinfo {volume} {84}},\ \bibinfo {pages} {052103} (\bibinfo {year} {2011})}\BibitemShut {NoStop}%
\bibitem [{\citenamefont {Nielsen}\ and\ \citenamefont {Chuang}(2010)}]{Nielsen_Book}%
  \BibitemOpen
  \bibfield  {author} {\bibinfo {author} {\bibfnamefont {M.}~\bibnamefont {Nielsen}}\ and\ \bibinfo {author} {\bibfnamefont {I.}~\bibnamefont {Chuang}},\ }\href {https://books.google.it/books?id=-s4DEy7o-a0C} {\emph {\bibinfo {title} {Quantum Computation and Quantum Information: 10th Anniversary Edition}}}\ (\bibinfo  {publisher} {Cambridge University Press},\ \bibinfo {year} {2010})\BibitemShut {NoStop}%
\bibitem [{\citenamefont {Simon}(2000)}]{Simon00}%
  \BibitemOpen
  \bibfield  {author} {\bibinfo {author} {\bibfnamefont {R.}~\bibnamefont {Simon}},\ }\bibfield  {title} {\bibinfo {title} {Peres-horodecki separability criterion for continuous variable systems},\ }\href {https://doi.org/10.1103/PhysRevLett.84.2726} {\bibfield  {journal} {\bibinfo  {journal} {Phys. Rev. Lett.}\ }\textbf {\bibinfo {volume} {{\bf{84}}}},\ \bibinfo {pages} {2726} (\bibinfo {year} {2000})}\BibitemShut {NoStop}%
\end{thebibliography}%


\begin{thebibliography}{1}

\bibitem{Weiss12}
U.~Weiss.
\newblock {\em Quantum Dissipative Systems}.
\newblock WORLD SCIENTIFIC, 4th edition, (2012).

\bibitem{Carrega22}
M.~Carrega, L.~M. Cangemi, G.~De~Filippis, V.~Cataudella, G.~Benenti, and M.~Sassetti.
\newblock Engineering dynamical couplings for quantum thermodynamic tasks.
\newblock {\em PRX Quantum}, {\bf 3}, 010323, (2022).

\bibitem{Cavaliere22}
F.~Cavaliere, M.~Carrega, G.~De~Filippis, V.~Cataudella, G.~Benenti, and M.~Sassetti.
\newblock Dynamical heat engines with non-markovian reservoirs.
\newblock {\em Phys. Rev. Res.}, {\bf 4}, 033233, (2022).

\bibitem{Cavaliere25}
F.~Cavaliere, G.~Gemme, G.~Benenti, D.~Ferraro, and M.~Sassetti.
\newblock Dynamical blockade of a reservoir for optimal performances of a quantum battery.
\newblock {\em Commun. Phys.}, {\bf 8}, 76, (2025).

\bibitem{Golub13}
G.~H. Golub and C.~F. van Loan.
\newblock {\em Matrix computations}.
\newblock John Hopkins University Press, 4th edition, (2013).

\end{thebibliography}
\clearpage
\onecolumngrid
\appendix
\section*{End Matter}    
\twocolumngrid

\setcounter{equation}{0}
\renewcommand{\thesection}{A}
{\em Thermodynamic efficiency}---We introduce here a figure of merit that accounts for the total thermodynamic cost of the charging process. It is based on the concept of entropy production, $
\Sigma_{\rm irr}(t) =\frac{\Delta E_{\mathrm{R}}(t)}{T}+\Delta S_{\mathrm{B}}(t)
$~\cite{Esposito10, Rao16, Landi21, Aguilar25} which, due to the second law of thermodynamics, is always $\ge 0$. 
The presence of a finite entropy production originates from two contributions: the heat exchanged with the reservoir,
$\Delta E_{\mathrm{R}}(t)=\langle H_{\mathrm{R}}(t) \rangle -  \langle H_{\mathrm{R}}(0) \rangle$, and to the Von Neumann entropy~\cite{Nielsen_Book} variation of the quantum battery, $\Delta S_{\mathrm{B}}(t)=S_{\mathrm{B}}(t)-S_{\mathrm{B}}(0)$. Since the total system is closed
while the battery is coupled to the reservoir, the total energy balance reads
$\Delta E_{\mathrm{R}}(t)+ \Delta E_{\mathrm{B}}(t)=W(t)$, which links the heat exchanged to  the work $W (t)=\langle H_{\mathrm{int}}(0) \rangle-\langle H_{\mathrm{int}}(t) \rangle$ done to switch on/off the coupling~\cite{SM} and to the accumulated energy in the battery,
$\Delta E_{\mathrm{B}}(t)=\langle H_{\mathrm{B}}(t) \rangle -  \langle H_{\mathrm{B}}(0)\rangle$. This allows us to rewrite the entropy production as 
\begin{equation}
\Sigma_{\rm irr}(t)=\frac{1}{T}\left[W(t)-\Delta E_{\mathrm{B}}(t) + T\Delta S_{\mathrm{B}}(t)\right ]\geq 0\,.
\label{entprod1}
\end{equation}

To define the efficiency associated to the global ergotropy, we rewrite $\Delta E_B(t)$ in terms of $\mathcal{E}_{\mathrm{glob}}(t)$. Using Eqs.~(\ref{eq:Etot}) and (\ref{eglob}), we have 
\begin{equation}
\!\Delta E_B(t)\!=\!\mathcal{E}_{\mathrm{glob}}(t)+\hbar\omega_0\left[
\nu_1(t)\!-\!\frac{C(T_0)}{2}\right]\!\equiv\mathcal{E}_{\mathrm{glob}}(t)-\Delta\epsilon(t).\nonumber
\end{equation}
The thermodynamic efficiency of the quantum battery can be now defined  using the concept of {\em exergy},  $\Phi(t)$ \cite{Manzano20, Lu23, Lopez23}, defined as the ratio between the negative
contributions to the entropy production (taken with a change of sign), $\Sigma_{irr}^{(-)}(t)$, and the positive ones $\Sigma_{\rm irr}^{(+)}(t)$:
$\Phi(t)={-\Sigma_{\rm irr}^{(-)}(t)}/{\Sigma_{\rm irr}^{(+)}(t)}$. Due to the second law, which guarantees $\Sigma_{\rm irr}(t)\ge 0$, the exergy is bounded as follows: $0\leq\Phi\leq 1$~\cite{Manzano20, Lu23, Lopez23}. 

Considering our system, we always observe a positive work required to switch on and off the interaction, i.e., $W(t) > 0$. Under this condition, the exergy $\Phi(t)$ can be written explicitly as:
\begin{equation}
\!\!\Phi(t)=\!\frac{\mathcal{E}_{\mathrm{glob}}(t)\!-\!\Delta\epsilon(t) 
\Theta[-\Delta\epsilon(t)]\!-\!T\Theta[-\Delta S_{\mathrm{B}}(t)]\Delta S_{\mathrm{B}}(t)
}{W(t) + \theta[\Delta\epsilon(t)]\Delta\epsilon(t) +
T\Theta[\Delta S_{\mathrm{B}}(t)]\Delta S_{\mathrm{B}}(t)}.\nonumber
\end{equation}
Now, in order to obtain a figure of merit associated with the ergotropy, we consider only the first term in the numerator of the exergy. This leads to the definition of the thermodynamic efficiency $\eta_{\rm th}(t)$, as reported in Eq.~(\ref{eq:etaTh}), which always satisfies 
$0\le\eta_{\rm th}\le\Phi\le 1$.
Note that for a Gaussian system, the Von Neumann entropy $S_B(t)$ of the quantum battery can be expressed in terms of the symplectic eigenvalues $\nu_l(t)$ of $\sigma_{\mathrm{B}}^{\rm glob}(t)$ as $S_B(t)= k_{\rm B}\sum^{N}_{j=1}f[\nu_{j}(t)]$ with 
$
f(\nu)=\sum_{p=\pm 1}p\left(\nu+\frac{p}{2} \right)\ln\left( \nu+\frac{p}{2}\right)$~\cite{Serafini2017}.
In our case, using the expression of $\nu_l(t)$ -- see Eq.(\ref{eq:nu1}) -- we obtain  $\Delta S_{\mathrm{B}}(t)=S_{\mathrm{B}}(t)-S_{\mathrm{B}}(0)=k_{\rm B}[f\left(\nu_1(t)\right)-f\left(\frac{C(T_0)}{2}\right)]$.\\

\setcounter{equation}{0}
\renewcommand{\thesection}{B}
{\em Squeezing}---The most general non--displaced single-mode Gaussian state is known to be a phase--space rotated {\em squeezed state}~\cite{Serafini2017,Brask2022}. As seen in Eq.~(\ref{eq:SigmaBM}), $\Sigma_{\mathrm{BM}}(t)$ is not diagonal and the presence of off--diagonal terms confirms this rotation. In general, the BM covariance matrix can be written as
\begin{equation}
\label{eq:gengauss}
{{\Sigma}}_{\mathrm{BM}}(t)={{R}}[\phi(t)]{{S}}[r(t)]\begin{pmatrix}\nu_1(t)&0\\0&\nu_1(t)\end{pmatrix}{{S}}[r(t)]^T{{R}}[\phi(t)]^T\nonumber\\\,,
\end{equation}
where $\nu_1(t)$ is its symplectic eigenvalue -- see Eq.~(\ref{eq:nu1}),
$
 {{R}}[\phi(t)]=\begin{pmatrix}
     \cos\left[\phi(t)\right] & \sin\left[\phi(t)\right]\\
     -\sin\left[\phi(t)\right] & \cos\left[\phi(t)\right]
 \end{pmatrix} $
is the $2\times2$ rotation matrix and ${{S}}[r(t)]=\mathrm{diag}\{e^{-r(t)},e^{r(t)}\}$ is the squeezing operator with squeezing parameter $r(t)$. 
Diagonalizing $\Sigma_{\mathrm{BM}}(t)$ amounts to expressing the state in terms of its principal axes, corresponding to a new set of canonical quadratures, $Q'_{\mathrm{BM}}(t)$ and $P'_{\mathrm{BM}}(t)$, along which the covariance matrix is diagonal. The variances along these axes are given by the eigenvalues of $\Sigma_{\mathrm{BM}}(t)$, namely $\lambda_{Q'_{\mathrm{BM}}}(t)=\langle {Q'}^2_{\mathrm{BM}}(t)\rangle$ and $\lambda_{P'_{\mathrm{BM}}}(t)=\langle {P'}^2_{\mathrm{BM}}(t)\rangle$. 
Incidentally, the angle between $Q'_{\mathrm{BM}}(t)$ and $Q_{\mathrm{BM}}(t)$ is precisely $\phi(t)$ with $\tan\left[2\phi(t)\right]=\frac{2b(t)}{a(t)-c(t)}$. The diagonal form of $\Sigma_{\mathrm{BM}}(t)$ is
\begin{equation}
\label{eq:gengauss}
\!{{\Sigma}}'_{\mathrm{BM}}(t)\!\!\equiv\!\!\begin{pmatrix}\lambda_{Q'_{\mathrm{BM}}}(t)&0\\0&\lambda_{P'_{\mathrm{BM}}}(t)\end{pmatrix}\!\!=\!\!\begin{pmatrix}e^{-2r(t)}\nu_1(t)&0\\0&e^{2r(t)}\nu_1(t)\end{pmatrix}\nonumber.
\end{equation}
From this one can immediately express the squeezing parameter as
$
r(t)=\frac{1}{4}\ln\left[\frac{\lambda_{P'_{\mathrm{BM}}}(t)}{\lambda_{Q'_{\mathrm{BM}}}(t)}\right]$. 
It is convenient to define $\lambda_+(t)$ (respectively, $\lambda_-(t)$) as the maximum (minimum) of the above eigenvalues:
$
\lambda_{\pm}(t) = \max/\min \left\{ \lambda_{Q'_{\mathrm{BM}}}(t), \lambda_{P'_{\mathrm{BM}}}(t) \right\}.
$
We have
\begin{equation}
\label{eq:eigenvaluespm}
\!\!\!\lambda_{\pm}(t)=\frac{1}{2}\left\{C(T_0)+\bar{\alpha}^2\!\left[\mathcal{T}(t)\pm\sqrt{\mathcal{T}^2(t)-4\Delta(t)}\right]\right\},
\end{equation}
obtained from Eq.~(\ref{eq:SigmaBM}), and satisfying
\begin{equation}
\lambda_{+}(t)\lambda_{-}(t)=\nu_1^2(t).
\label{legame}
\end{equation}

When $r(t)\neq 0$, squeezing occurs, and two qualitatively different regimes can be distinguished. When both $\lambda_{\pm}(t)>1/2$ one refers to {\em classical squeezing}: both variances are large and dominated by semiclassical fluctuations. In this case, squeezing merely reflects the anisotropy of the two variances. Conversely, when $\lambda_{-}(t)<\frac{1}{2}$ one of the two variances falls {\em below the ground state limit}, indicating a regime of {\em genuinely quantum squeezing}, whose origin cannot be ascribed to any classical effect.
\noindent Finally, recalling that the energy of the BM is expressed as $E_{\mathrm{BM}}(t)=\frac{\hbar\omega_0}{2}\mathrm{Tr}\{\Sigma_{\mathrm{BM}}(t)\}$, we obtain $E_{\mathrm{BM}}(t)=\frac{\hbar\omega_0}{2}\left[\lambda_{+}(t)+\lambda_{-}(t)\right]$, thus recovering Eq.~(\ref{eq:relationships}).\\

\setcounter{equation}{0}
\renewcommand{\thesection}{C}
{\em Logarithmic negativity}---In the following, we derive a closed expression for the logarithmic negativity $\mathcal{N}(t)$, which serves as a witness of entanglement~\cite{Serafini2017}. We consider a  balanced bipartition (subsystems A and B) of the $N$ (even) oscillators composing the QB, with $N_{\rm A}=N_{\mathrm{B}}=N/2$ and  $\sum_{l=1}^{N/2}\alpha^2_l=\sum_{l=1+N/2}^{N}\alpha^2_l=\bar\alpha^2/2$.
Since the dynamics of the QB is entirely captured by the BM, it is sufficient to consider a bipartition of the BM alone. This is done by introducing new quadratures $Q_{\mu}$, $P_{\mu}$ ($\mu=$A,B), corresponding to the two partitions,  defined as
$(Q_{\mathrm{A}},P_{\mathrm{A}})=\frac{\sqrt 2}{\bar{\alpha}}\sum_{j=1}^{N/2}\alpha_j(Q_j,P_j)$ and 
$(Q_{\mathrm{B}},P_{\mathrm{B}})=\frac{\sqrt 2}{\bar{\alpha}}\sum_{j=1+N/2}^{N}\alpha_j(Q_j,P_j)$. Given the bipartition state vector $(Q_{\mathrm{A}}(t),P_{\mathrm{A}}(t),Q_{\mathrm{B}}(t),P_{\mathrm{B}}(t))^T$ its covariance matrix can be written in block form as ${{\sigma}}_{\mathrm{bip}}(t)=\begin{pmatrix}{{L}}(t)&{{K}}(t)\\{{K}}^t(t) &{{L}}(t)\end{pmatrix}$.
These block matrices can be expressed in terms of the matrix elements of the covariance matrix. We have:
\begin{equation}
\label{eq:blockS1}
{{L}}(t)=\frac{C(T_0)}{2}\mathbbm{1}+{{K}}(t),\quad
{{K}}(t)=\frac{\bar\alpha^2}{2}\begin{pmatrix}
a(t) & b(t)\\
b(t) & c(t)
\end{pmatrix}\,.\nonumber
\end{equation}
The logarithmic negativity $\mathcal{N}(t)$ is then defined as
$
\mathcal{N}(t)=\sum_{j=\pm}\mathrm{max}\left\{0,-\ln\left[2\nu_j^{(\mathrm{PT})}(t)\right]\right\}$,
where $\nu_{-}^{(\mathrm{PT})}(t)\leq\nu_{+}^{(\mathrm{PT})}(t)$ are the symplectic eigenvalues of the {\em partially transposed covariance matrix} ${{\sigma}}_{\mathrm{bip}}^{(\mathrm{PT})}(t)={{\Lambda}}{{\sigma}}_{\mathrm{bip}}(t){{\Lambda}}$, where ${{\Lambda}}=\mathrm{diag}\{1,1,1,-1\}$ ~\cite{Serafini2017}. Entanglement is present if and only if $\mathcal{N}(t)>0$~\cite{Simon00}, which requires at least one symplectic eigenvalue $\nu_j^{(\mathrm{PT})}(t)<1/2$. 
However, since it is always $\nu_+^{(\mathrm{PT})}(t)\nu_-^{(\mathrm{PT})}(t)\geq1/4$, only $\nu_-^{(\mathrm{PT})}(t)$, can be smaller than 1/2 which leads to the final expression quoted in Eq.~(\ref{eq:logneg}):
\begin{equation}
\label{eq:negapp}
\mathcal{N}(t)=\mathrm{max}\left\{0,-\ln\left[2\nu_-^{(\mathrm{PT})}(t)\right]\right\}.
\end{equation}

To proceed, we evaluate $\nu_-^{(\mathrm{PT})}(t)$ from the explicit expression of ${{\sigma}}_{\mathrm{bip}}^{(\mathrm{PT})}(t)$, following standard procedures~\cite{Serafini2017}. After some algebra we find
\begin{equation}
\label{eq:numinusPTeqbip}
\!\nu_-^{(\mathrm{PT})}(t)\!=\!\frac{\sqrt{C(T_0)}}{2}\sqrt{C(T_0)+\bar{\alpha}^2\!\left[\mathcal{T}(t)\!-\!\sqrt{\mathcal{T}^2(t)\!-\!4\Delta(t)}\right]}.
\end{equation}
Comparing Eq.~(\ref{eq:numinusPTeqbip}) and Eq.~(\ref{eq:eigenvaluespm}), 
we immediately obtain the important relation
\begin{equation}
\nu_-^{(\mathrm{PT})}(t)=\sqrt{\frac{C(T_0)}{2}}\sqrt{\lambda_{-}(t)}\,,
\label{nuPT}
\end{equation}
quoted in Eq.~(\ref{eq:result1}). This equality allows us
to conclude that the presence of entanglement ($\nu_-^{(\mathrm{PT})}(t)<1/2$) directly implies $\lambda_{-}(t)<\frac{1}{2}$. Thus, for a balanced bipartition, the condition: Entanglement~$\implies$~Quantum squeezing is always fulfilled. Notice that the converse is in general not true.\\

\setcounter{equation}{0}
\renewcommand{\thesection}{D}
\noindent {\em Quantum advantage bounds}
---We compare here classical versus quantum regimes by choosing two configurations with the same number of oscillators $N$, equal temperatures $T_0$ and $T$, and same energy $E_{\mathrm{B}}(t)$ stored in the battery and, consequently, the same BM energy $E_{\mathrm{BM}}(t)$. Let us start by rewriting the global efficiency $\eta_{\rm glob}(t)$ in Eq.~(\ref{eq:etaGlob}), using the link derived in Eq.~(\ref{legame}). 
We have
\begin{equation}
\label{EM:etaGlob}
\eta_{\mathrm{glob}}(t)=1-\frac{\hbar\omega_0}{E'_{\mathrm{BM}}(t)}\left[\sqrt{\lambda_-(t)\lambda_+(t)}-1/2\right]\,.
\end{equation}
Now we express $\lambda_+(t)$
in terms of $\lambda_-(t)$ using the relation (\ref{eq:relationships}):
$
\lambda_{+}(t)=\frac{2E_{\mathrm{BM}}(t)}{\hbar\omega_0}-\lambda_{-}(t)$.
Inserting this expression  into $\eta_{\rm glob}(t)$, we write it in terms of $\lambda_-(t)$ only:
\begin{equation}
\!\!\!\eta_{\rm glob}(t)\!=\!1\!-\!\frac{1}{\kappa(t)}\left[\sqrt{\left[2\kappa(t)+1 -\lambda_-(t)\right]\lambda_-(t)}-1/2\right]\,,
\label{EM:etaglob}
\end{equation}
where $\kappa(t)=E'_{\mathrm{BM}}(t)/\hbar\omega_0>0$.
Notice that since $\lambda_+(t)\ge\lambda_-(t)$ it follows that $\lambda_-(t)\le \kappa(t)+1/2$. Therefor, $\eta_{\rm glob}(t)$ 
is monotonically decreasing in $\lambda_-(t)$. To obtain  the  explicit expression for the classical bound  
$\mathcal{B}_{\rm cl}(t)$, we impose $\lambda_-(t)=1/2$, corresponding to the boundary between classical and quantum squeezing, in Eq.~(\ref{EM:etaglob}). The result of $\mathcal{B}_{\rm cl}(t)=\eta_{\rm glob}(t)|_{\lambda_-(t)=1/2}$ is quoted in Eq.~(\ref{eq:bound1}).
Notice that, since $\eta_{\rm glob}(t)$ decreases  increasing  $\lambda_-(t)$, we can always conclude that if $\lambda_-(t)>1/2$ (classical squeezing) it is $0\leq\eta_{\mathrm{glob}}(t)<\mathcal{B}_{\mathrm{cl}}(t)$, on the other hand if $\lambda_-(t)<1/2$ (quantum squeezing) it is $\mathcal{B}_{\mathrm{cl}}(t)<\eta_{\mathrm{glob}}(t)\leq 1$. Moving from classical to  quantum regime is then possible by varying the  value of $\lambda_-(t)$. Notice that this procedure is achievable,  even if the energy $E_{\rm BM}(t)$ is fixed, by varying 
$\gamma_0$ and $\omega_{\mathrm{D}}$.

We conclude by addressing the  comparison within the regime of quantum squeezing ($\lambda_{-}(t)<1/2$), distinguishing between the presence ($\nu_-^{(\mathrm{PT})}(t)<1/2$)  or absence $(\nu_-^{(\mathrm{PT})}\ge 1/2)$ of entanglement. To find the corresponding bound  $\mathcal{B}_{\mathrm{en}}(t)$, discussed in Eq.~(\ref{eq:bound2}), we consider the relation obtained in Eq.~(\ref{nuPT}), which gives:
$\lambda_{-}(t)=2\left[\nu_-^{(\mathrm{PT})}(t)\right]^2/C(T_0)$. 
The explicit form  of  the bound is then obtained  imposing   the threshold for entanglement, that is, $\nu_-^{(\mathrm{PT})}(t)=1/2$, which corresponds to $\lambda_-=1/(2C(T_0))$, 
into Eq.~(\ref{EM:etaglob}). This yields $\mathcal{B}_{\rm en}(t)=\eta_{\rm glob}(t)|_{\lambda_-(t)=1/(2C(T_0))}$, whose result is quoted in Eq.~(\ref{eq:bound2}).
The different regions can now be identified by following arguments analogous to those previously discussed.
$\mathcal{B}_{\mathrm{cl}}(t)<\eta_{\mathrm{glob}}(t)\le\mathcal{B}_{\mathrm{en}}(t)$, in the regime with quantum squeezing without entanglement ($\frac{1}{2C(T_0)}\le\lambda_-<\frac{1}{2}$), and $\mathcal{B}_{\mathrm{en}}(t)<\eta_{\mathrm{glob}}(t)\le 1$ with entanglement ($\lambda_-<\frac{1}{2C(T_0)}$).
\end{document}


\title{{\em Supplemental material for}\\
Quantum advantage bounds for a multipartite gaussian battery}

\author{F. Cavaliere}
\affiliation{\genova}
\affiliation{\CNR}
\author{D. Ferraro}
\email{dario.ferraro@unige.it}
\affiliation{\genova}
\affiliation{\CNR}
\author{M. Carrega}
\affiliation{\CNR}
\author{G. Benenti}
\affiliation{\como}
\affiliation{\infn}
\author{M. Sassetti}
\affiliation{\genova}
\affiliation{\CNR}

\maketitle

\appendix
\setcounter{equation}{0}
\renewcommand{\thesection}{S1.}
\section{S1. Langevin equation and its solution}
\label{app:Langevin}

From the total Hamiltonian of the $N$-oscillator quantum battery (QB) coupled to a reservoir -- see Eqs.~(1),(2) in the main text -- one can follow standard procedures \cite{Weiss12,Carrega22,Cavaliere22,Cavaliere25} to derive a coupled set of Langevin equations of motion for the battery oscillator operators $q_l(t)$.
They read
\begin{equation} 
\ddot{q}_l(t) + \omega_0^2 q_l(t) +\alpha_l\int_0^t ds \gamma(t-s) \sum_{l'=1}^N\alpha_{l'}\dot{q}_{l'}(s) +\alpha_l\gamma(t)\sum_{l'=1}^N\alpha_{l'}q_{l'}(0)=\frac{\alpha_l}{m}\xi(t),
\end{equation}
with  damping kernel $\gamma(t)$, and  thermal noise operator $\xi(t)$ defined as 

\begin{eqnarray}
\gamma(t)&=&\frac{\theta(t)}{m}\sum_{k=1}^{\infty}\frac{c_k^2}{m_k\omega_k^2}\cos(\omega_k\,t)\,,\nonumber\\
\xi(t)&=&\sum_{k=1}^{\infty} c_k\left[x_k(0)\cos(\omega_k\,t)+\frac{\pi_k(0)}{m_k\,\omega_k}\sin(\omega_k\,t)\right].
\end{eqnarray}
\\
Notice that the noise  depends only on the reservoir operators $x_k(0),\pi_k(0)$ at initial time with the reservoir assumed to be in equilibrium at temperature $T$. This implies $\langle \xi(t)\rangle= 0$ with reservoir correlator~\cite{Weiss12}

\begin{equation}
\hspace{-0.4cm}\langle \xi(t)\xi(t')\rangle=\hbar\int_0^{\infty} \frac{d\omega}{\pi} J(\omega)\left\{\coth\left(\frac{\hbar\omega}{2k_{\mathrm{B}} T}\right)\cos\left[\omega (t-t')\right]-i\sin\left[\omega (t-t')\right]\right\},
    \label{xiaverage}
\end{equation}
written in terms of the spectral density \cite{Weiss12}
\begin{equation}
    J(\omega)=\frac{\pi}{2}\sum_{k=1}^{\infty}\frac{c_k^2}{m_k\omega_k}\delta(\omega-\omega_k).
\end{equation}\\
\noindent Applying the Laplace transform to the Langevin equation we get

 \begin{equation}
 [\lambda^2 + \omega^2_0 + \lambda \alpha_l^2 \tilde{\gamma}(\lambda)] \tilde{q}_l
   (\lambda) + \alpha_l\lambda\tilde{\gamma}(\lambda)\sum_{l'\not=l}\alpha_{l'}\tilde{q}_{l'} (\lambda)
   = \lambda q_l (0) + \dot{q}_l (0) + \frac{\alpha_l \tilde{\xi} (\lambda)}{m}.
   \end{equation}
   \\
   The above set of equations can be now written in the compact form $
    {{{\bm \chi}}}^{-1}(\lambda)\tilde{{\bf q}}(\lambda) = \tilde{{\bf y}}(\lambda)$, 
   with the  $N$-dimensional vector ${\bf \tilde q}^t(\lambda) = (\tilde q_1(\lambda),\dots , \tilde q_N(\lambda))$, the  external vector ${\bf \tilde y}^t(\lambda) = ( \lambda q_1 (0) + \dot{q}_1 (0) +\alpha_1 \tilde{\xi} (\lambda)/m,\dots ,\lambda q_N (0) + \dot{q}_N (0) +\alpha_N \tilde{\xi} (\lambda)/m )$, and the $N\times N$ inverse resolvent  matrix:
   
\begin{eqnarray}
 {{{\bm \chi}}}^{-1}(\lambda)=\left(\begin{array}{ccccc}
\Delta_1(\lambda) & \lambda\alpha_1\alpha_2\tilde{\gamma}(\lambda) &\dots & \lambda\alpha_1\alpha_N\tilde{\gamma}(\lambda)\\
\lambda\alpha_2\alpha_1\tilde{\gamma}(\lambda) &\Delta_2(\lambda) &\dots &\lambda\alpha_2\alpha_N\tilde{\gamma}(\lambda) \\
\dots &\dots &\dots  &\dots\\
\lambda\alpha_N\alpha_1\tilde{\gamma}(\lambda) &\lambda \alpha_N\alpha_2 \tilde{\gamma}(\lambda) &\dots &\Delta_N(\lambda) 
\end{array}\right),
\end{eqnarray}
\\
with diagonal elements $\Delta_l (\lambda) = \lambda^2 + \omega^2_0 + \lambda \alpha_l^2 \tilde{\gamma}
(\lambda)\,.$
Upon inversion one obtains the formal solution for the oscillator operators $\tilde{{\bf q}}(\lambda)$ in the Laplace domain:
\begin{equation}
\tilde{{\bf q}}(\lambda)= {{{\bm \chi}}}(\lambda)\tilde{{\bf y}}(\lambda),
\label{qlambda}
\end{equation}
with the explicit form of the matrix elements of ${{{\bm \chi}}}(\lambda)$ given by:

\begin{eqnarray}
{\chi}_{ll}(\lambda)&=&\frac{\lambda^2+\omega_0^2+\sum_{j \neq l}\alpha_j^2\lambda\tilde{\gamma}(\lambda)}{[\lambda^2+\omega_0^2][\lambda^2+\omega_0^2+\bar\alpha^2\lambda\tilde{\gamma}(\lambda)]}\\
{\chi}_{ll'}(\lambda)|_{l\not= l'}&=&\frac{-\alpha_l\alpha_{l'}\lambda
\tilde{\gamma}(\lambda)}{[\lambda^2+\omega_0^2][\lambda^2+\omega_0^2+\bar\alpha^2\lambda\tilde{\gamma}(\lambda)]}.
\label{chilambdall}
\end{eqnarray}
Here, we introduce the global interaction weight  

\begin{equation}
    \bar\alpha=\sqrt{\sum_{j=1}^N\alpha_j^2}.
    \label{baralpha}
\end{equation}
\\
From the solution in Eq.~(\ref{qlambda})  the behavior of the position operators in the time domain follows immediately. They naturally separate in a homogeneous part, depending on the initial conditions, and  in a thermal contribution due to the noise: 
$q_l (t) = q_{l, \mathrm{h}} (t) + q_{l,\mathrm{th}} (t)$, with
\begin{equation} q_{l,\mathrm{h}} (t) = \sum_{l'=1}^N [q_{l'} (0) \dot{\chi}_{l l'} (t) + \dot{q}_{l'} (0) \chi_{l l'} (t)],\quad
 q_{l,\mathrm{th}} (t) = \int_0^t \mathrm{d} s \frac{\xi (s)}{m}\sum_{l'=1}^N\alpha_{l'}\chi_{l,l'}(t - s).
 \label{th}
\end{equation}
Here, $\chi_{l l'} (t)$ are the components of the matrix elements given in Eq.~(\ref{chilambdall}) transformed back to the time domain, satisfying the boundary conditions $\chi_{l,l'}(0)=0,\quad\dot{\chi}_{l,l'}(0)=\delta_{l,l'}\quad\, \ddot{\chi}_{l,l'}(0)=0$.

\noindent We now proceed to write the explicit form of the resolvent matrix reservoir ${\bm \chi}(\lambda)$ in the case of the Drude damping, given by 
\begin{equation}
\gamma(t)=\gamma_0\omega_D\, e^{-\omega_D\,t}\,\theta(t),\quad \tilde\gamma(\lambda)=\frac{\gamma_0\omega_D}{\lambda+\omega_D},   
\end{equation}
where $\gamma_0$ is the damping strength and $\omega_D$ the cut-off frequency of the reservoir. Inserting  this form in Eq.~(\ref{chilambdall}) we obtain:

\begin{eqnarray}
{\chi}_{ll}(\lambda)&=&\frac{(\lambda^2+\omega_0^2)(\lambda+\omega_D)+\lambda\Omega_0^2\sum_{l'=1, l'\neq l}^N\alpha_{l'}^2}{\lambda^2+\omega_0^2}\,
{\prod_{i=3}^5 \frac{1}{(\lambda +\mu_i)}},\\ 
{\chi}_{ll'}(\lambda)|_{l\not=l'}&=&\frac{-\alpha_l\alpha_{l'}\lambda\Omega_0^2}{\lambda^2+\omega_0^2}\,{\prod_{i=3}^5 \frac{1}{(\lambda +\mu_i)}}\,,\qquad {\rm with}\qquad \Omega_0=\sqrt{\gamma_0\omega_D}.
\label{chilambdall1}
\end{eqnarray}
\\
Notice that $\chi_ {ll'}(\lambda)$ has five poles. Two of them are free, completely decoupled from the reservoir:  $\mu_{1,2}=\pm i \omega_0$ while the other three, $\mu_i$ with $i=3,4,5$, satisfy the equation:

\begin{equation}
(\omega_0^2+\mu_i^2)(\omega_D-\mu_i)-\mu_i\bar\alpha^2\Omega_0^2=0.
\end{equation}
\\
They depend on the coupling with the reservoir via the global interaction $\bar\alpha$ defined in Eq.~(\ref{baralpha}). With the above decomposition, the key expressions of the time dependent matrix elements $\chi_{ll'} (t)$ is obtained via an inverse Laplace transform:

\begin{eqnarray}
{\chi}_{ll'}(t)&=&\frac{\sin(\omega_0 t)}{\omega_0}\delta_{ll'}
+
\alpha_l\,\alpha_{l'}\,\Big[A(t)
-\frac{\sin(\omega_0 t)}{\bar\alpha^2\omega_0}\,
\Big],\nonumber\\
A(t)&=&\Omega_0^2\,\sum_{i=3}^5\mu_ie^{-\mu_i t}\left[
{\prod_{r=1;r\not= i}^5 \frac{1}{(\mu_r-\mu_i)}}\right].
\label{chilambdallt}
\end{eqnarray}
Notice that the function $A(t)$ does not explicitly depend on the local coupling $\alpha_l$, but only on the poles $\mu_{i}$.
In addition, due to the initial conditions on $\chi_{ll'}(t)$ we have $A(0)=0,\;\dot A(0)=\frac{1}{\bar\alpha^2},\; \ddot A(0)=0$.

\setcounter{equation}{0}
\renewcommand{\thesection}{S2.}
\section{S2. Local and global covariance matrix}
In this part we introduce the covariance matrix (CM), associated to the QB, using local and global approaches.
\subsection{S2A. Local covariance matrix}
\label{app:loccovmat}
We start by introducing the dimensionless quadrature operators
of the QB:
\begin{equation}
Q_l (t) = \sqrt{\frac{m \omega_0}{\hbar}} q_l (t);\quad P_l (t) = \sqrt{\frac{1}{{\hbar m\omega_0}}}   {p}_l (t)=\sqrt{\frac{m}{{\hbar  \omega_0}}}   \dot{q}_l (t). 
\end{equation}
The general form of the CM is
\begin{equation}
  {{\sigma}}_{\mathrm{B}}(t)=\left( {\begin{array}{ccccc}
    \sigma_{Q_{1}, Q_{1}}(t) & \sigma_{Q_{1}, P_{1}}(t) & \cdots &  \sigma_{Q_{1}, Q_{N}}(t) & \sigma_{Q_{1}, P_{N}}(t) \\
    \sigma_{P_{1}, Q_{1}}(t) & \sigma_{P_{1}, P_{1}}(t) & \cdots & \sigma_{P_{1}, Q_{N}}(t)& \sigma_{P_{1}, P_{N}}(t) \\
    \vdots & \vdots & \ddots & \vdots & \vdots \\
    \sigma_{Q_{N}, Q_{1}}(t)  & \sigma_{Q_{N}, P_{1}}(t) & \cdots & \sigma_{Q_{N}, Q_{N}}(t) & \sigma_{Q_{N}, P_{N}}(t)\\
    \sigma_{P_{N}, Q_{1}}(t)  & \sigma_{P_{N}, P_{1}}(t) & \cdots & \sigma_{P_{N}, Q_{N}}(t) & \sigma_{P_{N}, P_{N}}(t)\\
  \end{array} } \right)\,,
\label{sigmamatrix}
\end{equation}
\\
and depends on symmetrized correlation functions of the form $\sigma_{A,B}(t)=\frac{1}{2} \langle \left(A(t)B(t)+B(t)A(t) \right)\rangle -\langle A(t)\rangle\langle B(t)\rangle$.
We recall the equilibrium initial conditions of the $N$ oscillators, at temperature $T_0$, given by the density matrix: $e^{-H_{\mathrm{B}}/k_{\mathrm{B}} T_0}/Z_{\mathrm{B}}$. This  implies $\langle Q_l(t)\rangle =\langle P_l(t)\rangle = 0$, $[Q_l(t),P_l(t)]=i$, $\langle Q_l(0)P_l(0)+P_l(0)Q_l(0)\rangle =0$ and 
\begin{equation}
\langle Q^2_l(0)\rangle = \langle P^2_l(0)\rangle=\frac{C(T_0)}{2},\qquad {\rm with }\qquad C(T_0)=\coth{\left(\frac{\hbar\omega_0}{2k_{\mathrm{B}}T_0}\right)}.
\end{equation}
Following Eqs.~(\ref{th}) and taking into account that the averages between  homogeneous and thermal contributions are always zero, the CM elements decouple into a homogeneous and a thermal contribution:
\begin{equation}
\sigma_{A,B}(t)=\sigma_{A,B}^{(\mathrm{h})}(t)+
\sigma_{A,B}^{(\mathrm{th})}(t)\,.
\end{equation}
The calculation of the homogeneous part is straightforward and we quote here the main steps only. We start by inserting, in the averages, the homogeneous behaviors of $q_l(t)$ and $\dot q_l(t)$, written in Eq.~(\ref{th}):

\begin{eqnarray}
&&\sigma^{(\mathrm{h})}_{Q_l Q_{l'}}(t) =\frac{C(T_0)}{2}\sum_{j=1}^{N} \Big[\dot{\chi}_{l,j}(t)\dot{\chi}_{l',j}(t) + \omega^2_0\,\chi_{l,j}(t)\chi_{l'\!,j}(t)\Big],\nonumber \\
&&\sigma^{(\mathrm{h})}_{Q_lP_{l'}}(t)=\frac{C(T_0)}{2\omega_0}\sum_{j=1}^{N} \Big[\dot{\chi}_{l, j}(t) \ddot{\chi}_{l'\!, j}(t) 
+\omega_{0}^2\, \chi_{l,j}(t) \dot{\chi}_{l', j}(t)\Big],\nonumber\\
&&\sigma^{(\mathrm{h})}_{P_l,P_{l'}}(t)= \frac{C(T_0)}{2\omega^2_0}\sum_{j=1}^N\Big[\ddot{\chi}_{l,j}(t)\ddot{\chi}_{l'\!,j}(t) + \omega^2_0\,\dot{\chi}_{l,j}(t)\dot{\chi}_{l'\!,j}(t)\Big].
\end{eqnarray}
\noindent To reach a compact form it is sufficient to use the explicit time behavior of the matrix elements 
$\chi_{ll'}(t)$ of the resolvent, quoted in Eq.~(\ref{chilambdallt}). The final result is:
\begin{equation}
\sigma^{(\mathrm{h})}_{Q_l Q_{l'}}(t) =\frac{C(T_0)}{2}\,\delta_{ll'}+\,\alpha_l\,\alpha_{l'}\,a_{\rm h}(t),\quad
\sigma^{(\mathrm{h})}_{Q_lP_{l'}}(t)=\alpha_l\,\alpha_{l'}\, b_{\rm h}(t),\quad
\sigma^{(\mathrm{h})}_{P_l,P_{l'}}(t)= \frac{C(T_0)}{2}\delta_{ll'}+\,\alpha_l\,\alpha_{l'}\,c_{\rm h}(t).
\end{equation}
Here, we introduced the three functions $a_{\rm h}(t),\,b_{\rm h}(t),\,c_{\rm h}(t)$: 
\begin{eqnarray}
&&a_{\mathrm{h}}(t)=\frac{C(T_0)}{2}\Big[ \bar\alpha^2\Big(\dot A^2(t)+\omega_0^2\, A^2(t)\Big)-\frac{1}{\bar\alpha^2}
\Big],
\nonumber \\
&&b_{\mathrm{h}}(t)=\frac{C(T_0)}{2\omega_0}\,\bar\alpha^2\dot A(t)\Big[\ddot A(t)+\omega_0^2\, A(t)
\Big],\nonumber\\
&&c_{\mathrm{h}}(t)= \frac{C(T_0)}{2\omega_0^2}\Big[ \bar\alpha^2\Big(\ddot A^2(t)+\omega_0^2\, \dot A^2(t)\Big)-\frac{\omega_0^2}{\bar\alpha^2}
\Big].
\label{abch}
\end{eqnarray}
\\
We remind that the global interaction weight $\bar\alpha$ and the function  $A(t)$ are defined  in Eqs. (\ref{baralpha}),(\ref{chilambdallt}) respectively.\\

\noindent Finally, we evaluate the thermal contributions. As a starting point, we write the time dependent evolutions of $q_{l,\mathrm{th}}(t)$, and $\dot q_{l,\mathrm{th}}(t)$ in Eq.~(\ref{th}) in the compact form: 

\begin{equation}
    q_{l,\mathrm{th}}(t) = \int_0^t {\rm d} s \frac{\xi (s)}{m}\bar\chi_l(t - s),\quad
    \dot q_{l,\mathrm{th}} (t) =\int_0^t {\rm d} s \frac{\xi (s)}{m}\frac{d}{dt}\bar\chi_l(t - s),
 \label{eq:th1}
\end{equation}
\\
with
\begin{equation}
    \bar\chi_l(t)\equiv\sum_{l'=1}^N\alpha_{l'}\chi_{l,l'}(t).    
\end{equation}
\\
Inserting $\chi_{l,l'}(t)$ given in Eq.~(\ref{chilambdallt}), the above  resolvent function can be rewritten as:

\begin{equation}
\bar\chi_l(t)=\alpha_l\, \bar\alpha^2\, A(t)=\alpha_l\, \sum_{i=3}^5e^{-\mu_i t}\,\mu_i\Omega_N^2\,
\left[{\prod_{r=1;r\not= i}^5 \frac{1}{(\mu_r-\mu_i)}}\right]\equiv\alpha_l \sum_{i=3}^5e^{-\mu_i t}\,\bar\chi^{(i)},
\label{barchi}
\end{equation}
\\
where, in the next--to--last equality, we introduced the characteristic frequency

\begin{equation}
\Omega_N=\bar\alpha\,\sqrt{\gamma_0\omega_D}=\bar\alpha\Omega_0.
\end{equation}
\\
Inserting now Eq.~(\ref{eq:th1}) in the CM elements we have:

\begin{eqnarray}
\hspace{-1.4cm}&&\sigma^{(\mathrm{th})}_{Q_l Q_{l'}}(t)\!=\!\frac{\omega_0}{2\,m\hbar}\int_0^t \!ds_1\int_0^t\! ds_2\,
\bar\chi_l(t-s_1)\bar\chi_{l'}(t-s_2)\langle\{\xi(s_1),\xi(s_2)\}\rangle,\nonumber\\
\hspace{-1.4cm}&&\sigma^{(\mathrm{th})}_{Q_l P_{l'}}(t)\!=\!\frac{1}{2\,m\hbar}\int_0^t \!\!ds_1\!\int_0^t\! ds_2\,
\bar\chi_l(t-s_1)\Big[\frac{d}{d t}\bar\chi_{l'}(t-s_2)\Big]\langle\{\xi(s_1),\xi(s_2)\}\rangle,\nonumber\\
\hspace{-1.4cm}&&\sigma^{(\mathrm{th})}_{P_l P_{l'}}(t)\!=\!\frac{1}{2\hbar\,m\omega_0}\int_0^t \!\!ds_1\!\int_0^t\!\!ds_2\!
\Big[\frac{d}{d t}\bar\chi_l(t-s_1)\Big]\!\Big[\frac{d}{d t}\bar\chi_{l'}(t-s_2)\Big]\langle\{\xi(s_1),\xi(s_2)\}\rangle,
\end{eqnarray}
\\
with the symmetric reservoir correlator represented using the average in Eq.~(\ref{xiaverage}):

\begin{equation}
\frac{1}{2}\langle\{\xi(s_1),\xi(s_2)\}\rangle=\hbar\int_{-\infty}^{\infty} \frac{d\omega}{2\pi} J(\omega)\left\{\coth\left(\frac{\hbar\omega}{2k_{\mathrm{B}} T}\right)\cos\left[\omega (s_1-s_2)\right]\right\}.
\label{xiaverage1}
\end{equation}
\\
With this Fourier representation the two time integrals can be  easily evaluated, due to the exponential time behavior of $\bar\chi_l(t)$ in Eq.~(\ref{barchi}). The final result is:

\begin{equation}
\sigma^{(\mathrm{\mathrm{th}})}_{Q_l Q_{l'}}(t) =\alpha_l\alpha_{l'}\,a_{\mathrm{th}}(t),\qquad\sigma^{(\mathrm{th})}_{Q_l P_{l'}}(t) =\alpha_l\alpha_{l'}\,b_{\mathrm{th}}(t),\qquad\sigma^{(\mathrm{th})}_{P_l P_{l'}}(t) =\alpha_l\alpha_{l'}\,c_{\mathrm{th}}(t),    
\end{equation}
\\
with
\begin{eqnarray}
&&a_{\mathrm{th}}(t)=\frac{\omega_0}{m}
\int_{-\infty}^{\infty} \frac{d\omega}{2\pi} \left[J(\omega)\coth\left(\frac{\hbar\omega}{2k_{\mathrm{B}} T}\right)\right]\sum_{j,r=1}^3L_{j,r},
\nonumber \\
&&b_{\mathrm{th}}(t)=-\frac{1}{m}
\int_{-\infty}^{\infty} \frac{d\omega}{2\pi} \left[J(\omega)\coth\left(\frac{\hbar\omega}{2k_{\mathrm{B}} T}\right)\right]\sum_{j,r=1}^3\mu_j\,L_{j,r},\nonumber\\
&&c_{\mathrm{th}}(t)= \frac{1}{m\omega_0}
\int_{-\infty}^{\infty} \frac{d\omega}{2\pi} \left[J(\omega)\coth\left(\frac{\hbar\omega}{2k_{\mathrm{B}} T}\right)\right]\sum_{j,r=1}^3\mu_j\mu_r\,L_{j,r}\label{eq:abcth}\,,
\end{eqnarray}
with 
\begin{equation}
J(\omega)=m\gamma_0\omega_D^2\frac{\omega}{\omega^2+\omega_D^2}\,,   
\end{equation}
the Drude spectral density, and 
\begin{equation}
L_{j,r}=\frac{\bar\chi^{(j)}\bar\chi^{(r)}}{(\mu_r+i\omega)(\mu_j-i\omega)}\left[1+
e^{-(\mu_j+\mu_r)t}-e^{-(\mu_j-i\omega)t}-e^{-(\mu_r+i\omega)t}\right]\,.
\end{equation}
\\
We remind that the weights $\bar\chi^{(j)}$, associated to the poles $\mu_j$, are written in Eq.~(\ref{barchi}).\\

\noindent We conclude by summarizing the total expressions of the CM matrix elements. We have: 

\begin{equation}
\label{eq:sigmaloc}
\sigma_{\mathrm{B}}(t)=\frac{C(T_{0})}{2}\,  \mathbbm{1}+\left( {\begin{array}{cccc}
    \Sigma^{(1,1)}(t)&\cdots &\Sigma^{(1,N)}(t) \\
    \vdots & \ddots & \vdots\\
    \Sigma^{(N,1)}(t)&\cdots & \Sigma^{(N,N)}(t)\\
  \end{array} } \right)\,,
\end{equation}
where
\begin{equation}
\label{eq:sigmalocelementary}
 \Sigma^{(i,j)}(t) = \alpha_i\alpha_j\left( {\begin{array}{cc}
    a(t) & b(t)\\
    b(t) & c(t)\\
 \end{array} } \right)\,.
\end{equation}
is the block matrix due to the  QB/reservoir coupling with elements:
\begin{equation}
a(t)= a_{\mathrm{h}}(t)+a_{\mathrm{th}}(t),\qquad  
b(t)= b_{\mathrm{h}}(t) +b_{\mathrm{th}}(t),\qquad 
c(t)= c_{\mathrm{h}}(t)+c_{\mathrm{th}}(t)  
\end{equation}
\\
given in Eqs.~(\ref{abch}) and~(\ref{eq:abcth}).

\subsection{S2B. Global covariance matrix}
\label{app:DefBrightDark}
In the previous Section we described a {\em local} approach to the problem, with the CM written in terms of the quadratures of each individual oscillator. Here we outline the {\em global} approach, discussed in the main part, where the CM is expressed in terms of global coordinates.\\
\noindent All starts observing that the bright mode (BM), described by the quadratures   $Q_{\mathrm{BM}}(t)\equiv\frac{1}{\bar{\alpha}} \sum^{N}_{l=1}\alpha_{l}Q_{l}(t)$ and $P_{\mathrm{BM}}(t)\equiv\frac{1}{\bar{\alpha}} \sum^{N}_{l=1}\alpha_{l}P_{l}(t)$ -- see Eq.~(3) in the main text -- is the only one coupled to the reservoir and thus evolving non trivially, as shown by Eq.~(2) of the main text. There then exist $N-1$ {\em freely evolving} dark modes (DMs), with canonical quadratures $Q_{\mathrm{DM}}^{(j)}(t)$, $P_{\mathrm{DM}}^{(j)}(t)$ ($2\leq j\leq N$), spaning a degenerate subspace orthogonal to the BM. A convenient way to generate them is to introduce a unitary transformation $U$ acting on the local quadratures such that $\mathbf{R}'=U\mathbf{R}$, with
\begin{equation}
\label{eq:mapping}
\mathbf{R}=(Q_1,P_1,\ldots,Q_N,P_N)^T\,;\,\mathbf{R}'=(Q_{\mathrm{BM}},P_{\mathrm{BM}},Q_{\mathrm{DM}}^{(2)},P_{\mathrm{DM}}^{(2)},\ldots,Q_{\mathrm{DM}}^{(N)},P_{\mathrm{DM}}^{(N)})^T\,,
\end{equation}
where the time argument has been omitted here for compactness. Here,
\begin{equation}
\label{eq:Householder}
{{U}}={{\bar{U}}}\otimes\mathbbm{1}_{2\times2},\quad\mathrm{with}\quad{{\bar{U}}}=\mathbbm{1}_{N\times N}-2\frac{\mathbf{v} \mathbf{v}^t}{|\mathbf{v}|^2}\,,
\end{equation}
with ${{\bar{U}}}$ is a Householder reflector~\cite{Golub13}, with $\mathbf{v}=\hat{\mathbf{A}}-\mathbf{e}_1$,
$\hat{\mathbf{A}}=\frac{1}{\bar{\alpha}}(\alpha_1,\ldots,\alpha_N)^T$, $\mathbf{e}_1=(1,0,\ldots,0)^T$. Note that since $\frac{\mathbf{v} \mathbf{v}^T}{|\mathbf{v}|^2}$ is a projector, $\bar{U}$ and thus $U$ are orthogonal, hence also unitary. The matrix $U$ is also {\em symplectic}, satisfying $U\Omega U^T=\Omega$ where $\Omega=\bigoplus_{j=1}^{N}J$, with
\begin{equation}
\label{eq:J}
J=\begin{pmatrix}
    0 & 1\\ -1 & 0
\end{pmatrix}
\end{equation} 
the standard $2\times2$ symplectic matrix. The proof follows immediately by exploiting the orthogonality of $U$ and the definition of $\Omega$. The matrix ${{\bar{U}}}$ maps $\mathbf{e}_1$ to $\hat{\mathbf{A}}$ (the BM) and provides an orthonormal basis for the DMs. The symplectic lift ${{U}}={{\bar{U}}}\otimes\mathbbm{1}_{2\times2}$ ensures the correct interleaved order of quadratures.\\
\noindent The global operators are canonical, satisfying $[Q_{\mathrm{BM}}(t),P_{\mathrm{BM}}(t)]=[Q_{\mathrm{DM}}^{(j)}(t),P_{\mathrm{DM}}^{(j)}(t)]=i$, while all other commutators are zero. Noting that $H_{\mathrm{B}}$ of Eq.~(1) of the main text is written in terms of the local quadratures as $H_{\mathrm{B}}=\frac{\hbar\omega_0}{2}\mathbf{R}^{T}\cdot\mathbf{R}$, it is clear from the above unitary transformation that $H_{\mathrm{B}}=\frac{\hbar\omega_0}{2}\mathbf{R'}^{T}\cdot\left( UU^T\right)\mathbf{R'}^T\equiv \frac{\hbar\omega_0}{2}\mathbf{R'}^{T}\cdot\mathbf{R'}^T$, hence in the global picture the battery Hamiltonian can be decomposed as $H_{\mathrm{B}}=H_{\mathrm{BM}}+H_{\mathrm{DM}}$
where
\begin{equation}
\label{eq:Hdecomp}
H_{\mathrm{BM}}=\frac{\hbar\omega_0}{2}\left[P^2_{\mathrm{BM}}+Q^2_{\mathrm{BM}}\right]\,;\quad H_{\mathrm{DM}}=\frac{\hbar\omega_0}{2}\sum_{j=2}^{N}\left[{P_{\mathrm{DM}}^{(j)}}^2+{Q_{\mathrm{DM}}^{(j)}}^2\right]\quad\mathrm{with}\quad [H_{\mathrm{BM}},H_{\mathrm{DM}}]=0\,. 
\end{equation}
\noindent We are now in the position to derive the CM in the global basis, ${{\sigma}}_{\mathrm{B}}^{(\mathrm{glob})}(t)$. Its general form is:

\begin{equation}
  {{\sigma}}_{\mathrm{B}}^{(\rm glob)}(t)=\left( {\begin{array}{ccccc}
    \sigma_{Q_{\rm QM}, Q_{\rm QM}}(t) & \sigma_{Q_{\rm BM}, P_{\rm BM}}(t) & \cdots &  \sigma_{Q_{\rm BM}, Q_{\rm DM}^{(N)}}(t) & \sigma_{Q_{\rm BM}, P_{\rm DM}^{(N)}}(t) \\
    \sigma_{P_{\rm BM}, Q_{\rm BM}}(t) & \sigma_{P_{\rm BM}, P_{\rm BM}}(t) & \cdots & \sigma_{P_{\rm BM}, Q_{\rm DM}^{(N)}}(t)& \sigma_{P_{\rm BM}, P_{\rm DM}^{(N)}}(t) \\
    \vdots & \vdots & \ddots & \vdots & \vdots \\
    \sigma_{Q_{\rm DM}^{(N)}, Q_{\rm BM}}(t)  & \sigma_{Q_{\rm DM}^{(N)}, P_{\rm BM}}(t) & \cdots & \sigma_{Q_{\rm DM}^{(N)}, Q_{\rm DM}^{(N)}}(t) & \sigma_{Q_{\rm DM}^{(N)}, P_{\rm DM}^{(N)}}(t)\\
    \sigma_{P_{\rm DM}^{(N)},Q_{\rm BM}}(t) & \sigma_{P_{\rm DM}^{(N)}, P_{\rm BM}}(t) & \cdots & \sigma_{P_{\rm DM}^{(N)}, Q_{\rm DM}^{(N)}}(t) & \sigma_{P_{\rm DM}^{(N)}, P_{\rm DM}^{(N)}}(t)\\
  \end{array} } \right)\,.
\label{sigmamatrixglob}
\end{equation}
\noindent From Eq.~(\ref{eq:Hdecomp}), in the global picture the initial density matrix decomposes as 

\begin{equation}
\rho_{\mathrm{tot}}(0)=\frac{e^{-H_{\mathrm{BM}}/k_{\mathrm{B}} T_0}}{Z_{\mathrm{BM}}} \otimes \frac{e^{-H_{\mathrm{DM}}/k_{\mathrm{B}} T_0}}{Z_{\mathrm{DM}}} \otimes  \frac{e^{-H_{\mathrm{R}}/k_{\mathrm{B}} T}}{Z_{\mathrm{R}}}\,.
\end{equation}
\\
From this thermal condition at $t=0$ there are no correlations between the BM and any of the DMs, and the same holds also between two {\em different} DMs, this implies  ${{\sigma}}_{\mathrm{B}}^{(\rm glob)}(0)=\frac{C(T_0)}{2}\mathbbm{1}$. Since the DMs are decoupled from the reservoir they evolve freely and therefore their density matrix remains thermal at all times. In addition, they are also decoupled from the BM and this implies that the off--diagonal blocks of ${{\sigma}}_{\mathrm{B}}^{(\rm glob)}(t)$ are always null. Therefore we have 

\begin{equation} 
{{\sigma}}_{\mathrm{B}}^{(\mathrm{glob})}(t)={{\Sigma}}_{\mathrm{BM}}(t)\bigoplus_{j=2}^{N}{{\Sigma}}_{\mathrm{DM}}^{(j)}(t),
\quad {\rm with}\quad 
{{\Sigma}}_{\mathrm{DM}}^{(j)}(t)= {{\Sigma}}_{\mathrm{DM}}=\mathrm{diag}\left\{\frac{C(T_0)}{2},\frac{C(T_0)}{2}\right\}
\quad \forall j\,,
\end{equation}
\\
namely it is in block--diagonal form. 
Finally, the explicit form of the BM block, 
\begin{equation}
\label{eq:appsigmabm}
{{\Sigma}}_{\mathrm{BM}}(t)=\frac{1}{2}\begin{pmatrix} \langle \left\{Q_{\mathrm{BM}}(t), Q_{\mathrm{BM}}(t) \right\} \rangle& \langle \left\{Q_{\mathrm{BM}}(t), P_{\mathrm{BM}}(t) \right\} \rangle\\ \langle \left\{P_{\mathrm{BM}}(t), Q_{\mathrm{BM}}(t) \right\} \rangle&\langle \left\{P_{\mathrm{BM}}(t), P_{\mathrm{BM}}(t) \right\} \rangle\end{pmatrix},
\end{equation}
\\
is promptly obtained from the definition of the bright mode coordinates in terms of the local ones. From Eq.~(\ref{eq:sigmaloc}) the BM block is
\begin{equation}
{{\Sigma}}_{\mathrm{BM}}(t)=\begin{pmatrix}
\frac{C(T_0)}{2}+\bar{\alpha}^2a(t)&\bar{\alpha}^2b(t)\\
\bar\alpha^2b(t)&\frac{C(T_0)}{2}+\bar{\alpha}^2c(t)\label{eq:sigmaBMexplicit}
\end{pmatrix}\,.
\end{equation}
From Eq.~(\ref{eq:Hdecomp}) we immediately observe that $E_{\mathrm{B}}(t)=\frac{\hbar\omega_0}{2}\mathrm{Tr}\left\{\sigma_{\mathrm{B}}^{(\mathrm{glob})}(t)\right\}$ and exploiting the explicit expressions of $\Sigma_{\mathrm{BM}}(t)$ and $\Sigma_{\mathrm{DM}}$ one immediately finds that
\begin{equation}
E_{\mathrm{B}}(t)=E_{\mathrm{BM}}(t)+(N-1)\frac{\hbar\omega_0}{2}C(T_0)\quad\mathrm{with}\quad E_{\mathrm{BM}}(t)=\frac{\hbar\omega_0}{2}\left\{C(T_0)+\bar{\alpha}^2\left[a(t)+c(t)\right]\right\}\,,
\end{equation}
as quoted in Eq.~(5) of the main text.\\

\noindent By symplectic invariance, the symplectic spectrum $\{\nu_j(t)\}$ ($1\leq j\leq N$) is easily obtained from $\sigma_{\mathrm{B}}^{(\mathrm{glob})}(t)$. For the BM we identify $\nu_1(t)=\sqrt{\det\{\Sigma_{\mathrm{BM}}(t)\}}$ and obtain
\begin{equation}
\label{nuBM}
\nu_1(t)=\frac{1}{2}\sqrt{C^2(T_0)+2\bar{\alpha}^2\left[C(T_0)\mathcal{T}(t)+2\bar{\alpha}^2\Delta(t)\right]}\,,
\end{equation}
where $\mathcal{T}(t)=a(t)+c(t)$ and $\Delta(t)=a(t)c(t)-b^2(t)$. The DM blocks remain thermal, with 
\begin{equation}
\nu_{j\in[2,N]}=\sqrt{\det\{\Sigma_{\mathrm{DM}}\}}=¢\frac{C(T_0)}{2}\,.
\label{nuDM}
\end{equation}

\setcounter{equation}{0}
\renewcommand{\thesection}{S3.}
\section{S3. Local versus global efficiencies}
\label{app: etaloc}
With a multipartite battery different extraction protocols can be employed. The two extreme cases are a {\em local} one, with $N$ unitary detectors each addressing a single oscillator, and a {\em global} one with a single unitary detector acting on the whole battery. In the {\em local} case one has 
\be
\mathcal{E}_{\mathrm{loc}}(t)=E_{\mathrm{B}}(t)-\hbar\omega_0\sum_{l=1}^{N}\nu_1^{(l)}(t)\,,
\ee
where $\nu_1^{(l)}(t)$ is the symplectic eigenvalue of the matrix $\frac{C(T_0)}{2}\mathbbm{1}+{{\Sigma}}^{(l,l)}(t)$ (see Eq.~(\ref{eq:sigmaloc})). It is given by 
\begin{equation}
\nu_{1}^{(l)}(t)=\frac{1}{2}\sqrt{C^2(T_0)+2\alpha_l^{2}\left[C(T_0)\,\mathcal{T}(t)+2 \alpha_l^{2} 
\Delta(t)\right]}\,.
\label{nulocal}
\end{equation}

\begin{figure}[h!t]
    \centering
\includegraphics[width=0.90\textwidth]{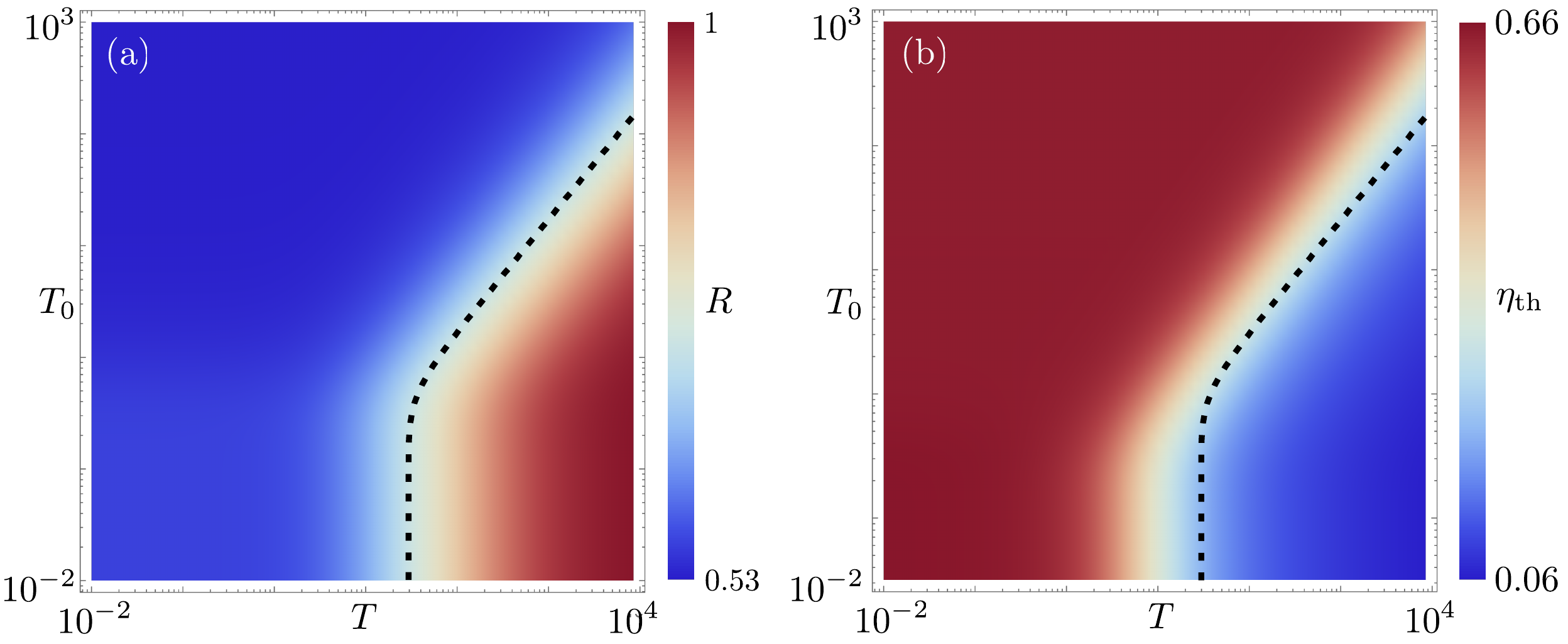}
    \caption{Density plot  as a function of $T$ and $T_0$ (units $\hbar\omega_0/k_{\mathrm{B}}$) for  (a) $R=R(t^*)$, and for (b) $\eta_{\mathrm{th}}=\eta_{\mathrm{th}}(t^*)$.  The black dashed line in both panels  represents the boundary $T=T^{*}=\frac{\hbar\Omega_N^2}{2k_{\mathrm{B}}T}\coth\left(\frac{\hbar\omega_0}{2k_{B}T_0}\right)$.
    The other parameters are $N=6$, $\alpha_l=1$, $\gamma_0=5\,\omega_0$ and $\omega_D=2\,\omega_0$.}
    \label{Supp:Fig1}
\end{figure}

For a global extraction the ergotropy is instead written in terms of the symplectic eigenvalues $\nu_j(t)$ of 
${{\sigma}}_{\mathrm{B}}^{(\mathrm{glob})}(t)$. We have:
\begin{equation}
\mathcal{E}_{\mathrm{glob}}(t)=E_{\mathrm{B}}(t)-\hbar\omega_0\sum_{j=1}^N\nu_j(t).
\end{equation}
From the expressions (\ref{nuBM}) and  (\ref{nuDM})
we easily arrive to the compact form:
\begin{equation}
\mathcal{E}_{\mathrm{glob}}(t)=E_{\mathrm{BM}}(t)-\hbar\omega_0\nu_1(t)\,.   
\label{eglob}
\end{equation}
We are now in the position to assess the most favorable extraction protocol. For this task, we analyze the behavior of the ratio $R(t)=\mathcal{E}_{\mathrm{loc}}(t)/\mathcal{E}_{\mathrm{glob}}(t)$.  A plot of $R(t^*)$ is shown in Fig.~\ref{Supp:Fig1}(a) as a function of $T$ and $T_0$. One observes two regions with different values, separated by the characteristic temperature $T^*=\frac{\hbar\Omega_N^2}{2k_{\mathrm{B}}T}\coth\left(\frac{\hbar\omega_0}{2k_{B}T_0}\right)$, emerging from the competition between the reservoir thermal fluctuations and those  of the initial condition of the QB. When $T>T^*$ (red region) thermal fluctuations  are large and they are not useful to enhance the global extraction since they only induce classical correlations, then  $R\sim 1$. On the other hand, when $T<T^*$ (blue region) thermal fluctuations are suppressed  allowing to extract more work via global operations, $R<1$.
For comparison we reproduce in panel (b)  the thermal efficiency $\eta_{\mathrm{th}}(t^*)$ defined in Eq. (9) of the main part and already shown there in Fig. 1(c). As we can see the behavior of  $R(t^*)$ is opposite to that of $\eta_{\mathrm{th}}(t^*)$. Indeed, for $T > T^*$, the efficiency  $\eta_{\mathrm{th}}(t^*)$ is strongly reduced  since thermal fluctuations impose a significant entropic cost.
We then conclude  that  in order to optimize the thermal cost  the best protocol of extraction is the  global one at $T<T^*$.
\setcounter{equation}{0}
\renewcommand{\thesection}{S4.}
\section{S4. Energy cost to switch on/off the coupling}
\label{app:work}
Switching on (at time $t=0$) and off (at time $t$) the coupling term $H_{\mathrm{int}}$ between the QB and the thermal reservoir requires performing work, thereby implying an energy cost. Here, we derive the closed-form expression for this work, defined as:

\begin{equation}
W(t)= \langle H_{\mathrm{int}}(0) \rangle-\langle H_{\mathrm{int}}(t) \rangle.
\end{equation}
\\
We start by recalling the interaction Hamiltonian
\begin{equation}
H_{\mathrm{int}}=-\sum^{+\infty}_{k=1}c_{k}x_{k} \sum^{N}_{l=1} \alpha_{l} q_{l}
+\sum^{+\infty}_{k=1}\frac{c^{2}_{k}}{2 m_{k} \omega^{2}_{k}}\left(\sum^{N}_{l=1} \alpha_{l} q_{l} \right)^{2}.
\label{hinteraction}
\end{equation}
\\
The switching on term at $t=0$ has a simple expression:
\begin{equation}
\langle H_{\mathrm{int}}(0) \rangle =
\sum^{+\infty}_{k=1}\frac{c^{2}_{k}}{2 m_{k} \omega^{2}_{k}}\sum^{N}_{l=1} \alpha_{l}^2 \langle q^2_{l}(0)\rangle=
\frac{\hbar\,\bar\alpha^2\Omega_0^2}{4\omega_0}\,C(T_0),
\end{equation}
where we have exploited the initial thermal average
$\langle q^2_{l}(0)\rangle=\frac{\hbar C(T_0)}{2m\omega_0}$ and we recall the definition 
\begin{equation}
\Omega_0^2=\sum^{+\infty}_{k=1}\frac{c^{2}_{k}}{m\,m_{k} \omega^{2}_{k}}=\gamma_0\omega_D
\end{equation}
\\
In order to express the switching off work at time $t$, we use the following relation, valid  for any $l$:
\begin{equation}
\alpha_l\sum^{+\infty}_{k=1}c_{k}x_{k}(t)=
m\left[\ddot q_l(t) +\omega_0^2\,q_l(t)+\Omega_0^2\alpha_l\sum^{N}_{l'=1} \alpha_{l'} q_{l'}(t)\right],
\label{relHint}
\end{equation}
\\
obtained from the equation of motion for the battery oscillators \cite{Cavaliere25}.
Inserting this expression into the general form of Eq.~(\ref{hinteraction})
we arrive at the following average:

\begin{equation}
\langle H_{\mathrm{int}}(t)\rangle=-m\sum^{N}_{l=1}\left[\frac{1}{2}\langle \{q_l(t),\ddot q_{l}(t)\}\rangle+\omega_0^2
\langle q^2_l(t)\rangle
-\frac{\Omega_0^2}{4}\sum^{N}_{l'=1}
\alpha_l\alpha_{l'}\langle \{q_l(t), q_{l'}(t)\}\rangle
\right] .
\end{equation}
\\
As a final step, we express the above relation in terms of the CM elements, using $\langle \{q_l(t),\ddot q_{l}(t)\}\rangle= \frac{d^2}{dt^2}\langle q_l^2(t)\rangle-2\langle \dot q_l^2(t)\rangle$. The final expression is

\begin{equation}
    \langle H_{\mathrm{int}}(t) \rangle= -\hbar \bar{\alpha}^{2} \left\{\frac{\ddot{a}(t)}{2 \omega_{0}} - \omega_{0} \left[c(t)-a(t) \right]+\frac{\Omega^{2}_{0}}{4 \omega_{0}}\left[C(T_0)+2 \bar{\alpha}^{2} a(t) \right]\right\}.
\end{equation}

\bibliography{references}
